\documentclass[10pt,aps,rmp,twocolumn,groupedaddress,longbibliography]{revtex4-1}
\usepackage{epsfig,amsmath,amssymb,verbatim,MnSymbol,bbm,dutchcal,mathtools}
\usepackage[colorlinks=true,citecolor=blue]{hyperref}

\begin{document}


\title{Coherent Transport in Periodically Driven Mesoscopic Conductors:
\mbox{From Scattering Matrices to Quantum Thermodynamics}}
\author{Kay Brandner}
\affiliation{
Department of Physics,
Keio University,
3-14-1 Hiyoshi, Kohoku-ku, 
Yokohama 223-8522,
Japan
\vspace*{6pt}}

\affiliation{
School of Physics and Astronomy,
University of Nottingham,
Nottingham NG7 2RD,
United Kingdom
\vspace*{6pt}}

\affiliation{Centre for the Mathematics and Theoretical Physics of Quantum Non-equilibrium Systems,
University of Nottingham,
Nottingham NG7 2RD, 
United Kingdom}




\begin{abstract}
Scattering theory is a standard tool for the description of transport
phenomena in mesoscopic systems. 
Here, we provide a detailed derivation of this method for nano-scale 
conductors that are driven by oscillating electric or magnetic fields.
Our approach is based on an extension of the conventional 
Lippmann-Schwinger formalism to systems with a periodically 
time dependent Hamiltonian. 
As a key result, we obtain a systematic perturbation scheme for the 
Floquet scattering amplitudes that describe the transition of a 
transport carrier through a periodically driven sample. 
Within a general multi-terminal setup, we derive microscopic
expressions for the mean values and time-integrated correlation 
functions, or zero-frequency noise, of matter and energy currents, 
thus unifying the results of earlier studies. 
We show that this framework is inherently consistent with the first 
and the second law of thermodynamics and prove that the mean rate of
entropy production vanishes only if all currents in the system are
zero. 
As an application, we derive a generalized Green-Kubo relation, which
makes it possible to express the response of any mean currents to
small variations of temperature and chemical potential gradients in 
terms of time integrated correlation functions between properly 
chosen currents. 
Finally, we discuss potential topics for future studies and further
reaching applications of the Floquet scattering approach to quantum 
transport in stochastic and quantum thermodynamics.
\end{abstract}


\maketitle

\vbadness=10000
\hbadness=10000

\newcommand{\braket}[2]{\langle #1|#2\rangle}
\newcommand{\bbraket}[2]{\bigl\langle #1\bigr| #2\bigr\rangle}
\newcommand{\BraKet}[2]{\llangle #1|#2\rrangle}
\newcommand{\negphantom}[1]{\settowidth{\dimen0}{#1}\hspace*{-\dimen0}}
\newcommand{\sint}[2]{\sumint\!d#2\negphantom{$\!d#2$}\hspace*{-4.5pt}\raisebox{-7pt}{{{\scriptsize $#1$}}}
\negphantom{{{\scriptsize $#1$}}}\phantom{d#2\hspace*{4pt}}}

\newcommand{\ket}[1]{|#1\rangle}
\newcommand{\bket}[1]{\bigl|#1\bigr\rangle}
\newcommand{\Ket}[1]{|#1\rrangle}

\newcommand{\bra}[1]{\langle #1 |}
\newcommand{\bbra}[1]{\bigl\langle #1 \bigr|}
\newcommand{\Bra}[1]{\llangle #1|}
\newcommand{\ix}[1]{\raisebox{-1.3pt}{{{\scriptsize $#1$}}}}
\newcommand{\dbtilde}[1]{\tilde{\raisebox{0pt}[0.85\height]{$\tilde{#1}$}}}

\renewcommand{\a}{\alpha}
\renewcommand{\b}{\beta}
\renewcommand{\d}{\delta}
\renewcommand{\l}{\lambda}
\newcommand{\g}{\gamma}
\newcommand{\ve}{\varepsilon}
\newcommand{\vp}{\varphi}
\newcommand{\TR}{\Theta}

\renewcommand{\tint}{\int_0^\tau \!\!\! dt\;}
\newcommand{\tpint}{\int_0^t \!\!\! dt'\;}
\newcommand{\tppint}{\int_0^t \!\!\! dt'' \;}
\newcommand{\Eint}{\int_0^\infty \!\!\! dE \;}
\newcommand{\Epint}{\int_0^\infty \!\!\! dE' \;}
\newcommand{\asum}{\sum\nolimits_\a}
\newcommand{\bsum}{\sum\nolimits_\b}
\newcommand{\gsum}{\sum\nolimits_\g}
\newcommand{\msum}{\sum\nolimits_m}
\newcommand{\nsum}{\sum\nolimits_n}
\newcommand{\psum}{\sum\nolimits_p}

\newcommand{\Em}{E\ix{m}}
\newcommand{\En}{E\ix{n}}
\newcommand{\Ep}{E\ix{p}}
\newcommand{\Fm}{E'\!\ix{m}}
\newcommand{\Fn}{E'\!\ix{n}}
\newcommand{\Fp}{E'\!\ix{p}}

\newcommand{\E}{\varepsilon}
\newcommand{\Jh}{\hat{\Jf}}
\newcommand{\Jf}{\mathsf{J}}
\newcommand{\If}{\mathsf{I}}
\newcommand{\Uf}{\mathsf{U}}
\newcommand{\Ac}{\mathcal{W}}
\newcommand{\Rc}{\mathcal{R}}
\newcommand{\sPhi}{\mathsf{\Phi}}
\newcommand{\jc}{k}
\newcommand{\jf}{\mathsf{j}}

\newcommand{\limtt}{\lim_{t\rightarrow\infty}\frac{1}{t}\!\int_0^t\!\!\! dt'\!\!\int_0^t\!\!\! dt''\;}
\newcommand{\limt}{\lim_{t\rightarrow\infty}\frac{1}{t}\!\int_0^t\!\!\! dt'\;}

\section{Introduction}
At room temperature, transport in macroscopic systems is a stochastic 
process, where carriers undergo ceaseless collisions that randomly 
change their velocity and direction of motion.
This irregular behavior is the microscopic origin of both the finite
resistance of a normal conductor and the fluctuations of induced
currents.
The fundamental relationship between these two phenomena is described
by the fluctuation-dissipation theorem, a cornerstone result of 
statistical mechanics, which goes back to the pioneering 
works of Einstein, Nyquist and Onsager and was later derived in a 
unified manner by Callen and Welton. 
Green and Kubo further expanded this approach and showed that, close 
to equilibrium, linear transport coefficients, which describe the 
response of a system to a small external field or thermal 
perturbation, can be expressed in terms of time integrated correlation
functions of the corresponding currents, i.e., the zero-frequency 
noise \cite{Kubo1966,Kubo1998,Marconi2008a}. 
This universal structure can be recovered even for systems in 
non-equilibrium steady states by introducing more general correlation
functions that involve a current and a suitably chosen conjugate 
variable \cite{Seifert2010}. 

Reducing the temperature of a conductor increases the average distance
that carriers can travel between two consecutive collisions. 
Coherent transport sets in when this mean free path becomes comparable
to the dimensions of the sample. 
In this regime, which is realized in mesoscopic systems at millikelvin
temperatures, the transfer of carriers becomes a reversible process 
governed by Schr\"odinger's equation. 
As a result, the properties of mesoscopic conductors are dominated by
quantum effects such as conductance quantization or coherent 
resistance oscillations, which can no longer be understood in terms of
classical stochastic trajectories \cite{Mello2004,Nazarov2009b,
Lesovik2014}.

Scattering theory provides a quantum mechanical description of open 
systems that are subject to a constant in- and outflow of particles . 
Therefore, it is a well suited tool to explore the principles of
coherent transport.
This approach was first proposed by Landauer and has since then 
evolved into a powerful theoretical framework, which has been 
extensively tested in experiments and shaped our modern 
understanding of transport phenomena in small-scale conductors. 
At the core of this framework lies the Landauer-B\"uttiker formula. 
It connects the scattering amplitudes of a mesoscopic sample, which 
describe the elastic deflection of incoming carriers, with the matter
and energy currents that emerge in the system under external biases. 
Hence, it provides a direct link between microscopic and macroscopic
quantities \cite{Buttiker1985,Buttiker1992,Blanter2000,Mello2004,
Nazarov2009b,Lesovik2014}. 

As a key application, the scattering approach to quantum transport
enables systematic investigations of the elementary principles that 
govern the thermodynamics of mesoscopic conductors and the performance
of autonomous nano-machines such as thermoelectric heat engines or 
refrigerators \cite{Gaspard2013a,Gaspard2015a,Gaspard2015,Benenti2017}.
Cyclic machines like charge pumps or quantum motors, however, require
the input or extraction of mechanical work;
therefore, they must be driven by time dependent electric or magnetic
fields, which alter the energy of carriers inside the sample. 
Floquet theory provides an elegant way to take this effect into 
account by introducing a new type of scattering amplitudes that 
describe inelastic transitions, where carriers exchange photons
with the external fields.
This Floquet scattering approach yields a generalized 
Landauer-B\"uttiker formula for periodically driven systems
\cite{Pedersen1998,Moskalets2002,Moskalets2012}.
Among other applications, this result makes it possible to develop
quantitative models for cyclic nano-machines, which can be used to 
engineer practical devices or to explore fundamental performance 
limits, two central quests in the field of quantum thermodynamics 
\cite{Vinjanampathy2016,Ludovico2016b}. 

As a second key result, the Floquet scattering approach leads to 
explicit microscopic expressions for the time integrated correlation 
functions of matter and energy currents in periodically driven quantum
conductors \cite{Moskalets2002a,Moskalets2004,Moskalets2014}. 
It thus provides an excellent basis to further investigate the complex
interplay between dissipation, thermal and quantum fluctuations in
mesocopic systems. 
This topic includes the search for generalizations of the 
well-established Green-Kubo relations as well as the quest for quantum
extensions of the recently discovered thermodynamic uncertainty 
relation \cite{Barato2015,Gingrich2016}. 

\section{Objective and Outline}
Our aim is to provide a thorough and general derivation of the Floquet
scattering approach to coherent transport in mesoscopic conductors. 
This article is supposed to serve as both a step-by-step introduction
for new users of the formalism and a compact reference text for 
experts in the field.  
We do not attempt to give a complete overview of the existing
literature. 
Instead, our objective is to complement earlier works by focusing on 
the development of an algebraic scattering theory for periodically 
driven systems and applications in stochastic and quantum
thermodynamics. 

We proceed as follows. 
In Sec.~\ref{Sec_MultTermMod}, we set the stage for our analysis by
introducing the multi-terminal model as a general basis for the 
discussion of coherent transport. 
This section is followed by a brief recap of the algebraic scattering
theory for autonomous systems in Sec.~\ref{Sec_FST_SC}, which is based
on common textbooks \cite{Newton1982,Ballentine1998,Mello2004,
Nazarov2009b}.
We then show how the Floquet theorem makes it possible to extend this
framework to periodically driven systems in Sec.~\ref{Sec_FST_DC}. 
Using the construction of an extended Hilbert space, which was 
originally proposed for closed systems \cite{Sambe1973}, we derive a 
generalized Lippmann-Schwinger equation for Floquet scattering states. 
This result naturally leads to a systematic perturbation scheme for
the crucial Floquet scattering amplitudes and to explicit expressions
for the corresponding scattering wave functions, which enable a 
transparent physical interpretation of the formalism. 

In Sec.~\ref{Sec_MaEnCurr} we switch from the single-particle picture
that had been used in the foregoing sections to a many-particle 
description. 
To this end we first show how the operators $\Jf^\rho_x$ and 
$\Jf^\ve_x$, which represent the matter and energy currents in a 
multi-terminal conductor, can be connected to the previously 
discussed Floquet scattering states. 
We then derive microscopic expressions for the mean currents and
the time integrated current correlation functions, or noise power,
which are given by 
\begin{subequations}\label{Int}
\begin{align}
\label{Int_CurrCorr}
J^x_\a &\equiv \lim_{t\rightarrow\infty}\frac{1}{t}\tpint
	\bigl\langle \Jf^{x\a}_t\bigr\rangle \quad\text{and}\\[6pt]
P^{xy}_{\a\b} &\equiv \lim_{t\rightarrow\infty} \frac{1}{t}
	\tpint\!\!\!\tppint \bigl\langle 
	(\Jf^{x\a}_{t'}- J^x_\a)(\Jf^{y\b}_{t''}- J^y_\b)\bigr\rangle,
\end{align}
\end{subequations} 
where angular brackets denote the average over all possible quantum
sates of the system. 
We thereby recover the results of earlier studies
\cite{Moskalets2002,Moskalets2004,Moskalets2014}. 

Moving on, in Sec.~\ref{Sec_Thermo} we show how the Floquet scattering
approach can be furnished with a thermodynamic structure. 
To this end, we formulate the first and the second law and show that 
the scattering formalism is inherently consistent with these 
constraints. 
As an application of this theory, we derive a generalization of the 
Green-Kubo relations for periodically driven systems far from 
equilibrium.
Finally, we discuss open problems and potential starting points for 
future studies in Sec.~\eqref{Sec_Open}. 

\section{The Multi-Terminal Model}\label{Sec_MultTermMod}

\begin{figure}[h!]
\begin{center}
\includegraphics[scale=1.055]{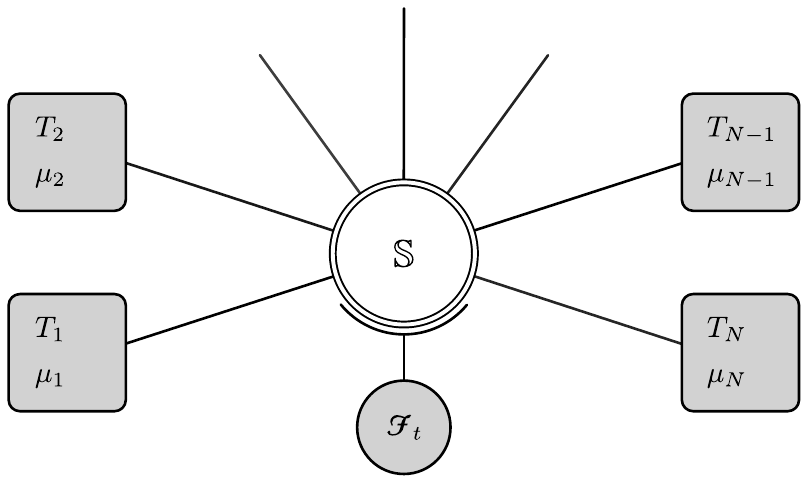}
\end{center}
\caption{Sketch of the multi-terminal model for a generic mesoscopic 
conductor. 
A central scattering region, or sample, $\mathbb{S}$ is connected via
ideal, one-dimensional leads to $N$ heat and particle reservoirs with 
temperatures $T_1,\dots,T_N$ and chemical potentials $\mu_1,\dots,
\mu_N$. 
The external driving fields $\boldsymbol{\mathcal{F}}_t$ periodically
change the potential inside the scattering region. 
\label{Fig_1}}
\end{figure}

The multi-terminal model provides a universal platform for the 
description of coherent transport in mesoscopic systems. 
The key idea is thereby to divide the conductor into a scattering
region, where carriers are affected by the potential landscape of the
sample and periodic driving fields, and a set of $N$ ideal leads, 
which can be traversed freely Fig.~\ref{Fig_1}.
For the sake of simplicity, we assume throughout this article that 
the leads are effectively one-dimensional.\footnote{Specifically,
we assume that the waveguides are so narrow that only the lowest 
transverse modes contribute to the transport process in the relevant
regime of energies, for details see \cite{Nazarov2009b}.}.

Each lead is connected to a thermochemical reservoir with fully 
transparent interface, which injects a continuous beam of thermalized,
non-interacting carriers into the system. 
Inside the conductor, these carriers follow a deterministic time 
evolution governed by Schr\"odingers equation until they are absorbed
again into one of the reservoirs.
Hence, all irreversible processes are relegated to the reservoirs, 
while the transfer of carriers between them is coherent. 
Once the system has reached a steady state, each lead $\a$ is 
traversed by a periodically modulated beam of incoming and outgoing
carries, which gives rise to a matter and an energy current. 
The corresponding mean values and fluctuations are given by the 
formulas \eqref{Int}. 
As we will see in the following sections, these quantities are 
completely determined by the scattering amplitudes of the driven 
sample and the energy distribution of the carriers injected by the 
reservoirs. 

\section{Standard Scattering Theory}\label{Sec_FST_SC}
\subsection{Scattering States}
Without external driving, the carrier dynamics in a multi-terminal 
system is governed by the Hamiltonian 
\begin{equation}\label{IIA1_Hamiltonian}
H = P^2/2M + U.
\end{equation}
Here $P$ and $M$ are the carrier momentum and mass and $U$ accounts
for the potential landscape of the scattering region as well as the 
coupling to external magnetic fields. 
The scattering of individual carriers with fixed energy $E>0$ is  
described by solutions of the time dependent Schr\"odinger equation 
that have the form 
\begin{equation}
\ket{\psi^{\a\pm}_{E,t}} = \exp[-i E t/\hbar]\ket{\vp^{\a\pm}_E}. 
\end{equation}
The outgoing and incoming states, $\ket{\vp^{\a+}_E}$ and 
$\ket{\vp^{\a-}_E}$, thereby represent carriers that enter and 
escape the system through the terminal $\a$, respectively. 
These scattering states satisfy the stationary Schr\"odinger equation 
\begin{equation}\label{IIA1_SSE}
H\ket{\vp^{\a\pm}_E} = E\ket{\vp^{\a\pm}_E}
\end{equation}
and the boundary conditions 
\begin{equation}\label{IIA1_SBC}
\braket{r_\b}{\vp^{\a\pm}_E} \equiv \vp^{\a\pm}_E[r_\b]
	=\d_{\a\b} w^\mp_E [r_\b]+S^{\a\b\pm}_E w^\pm_E[r_\b]
\end{equation} 
Here, the plane waves 
\begin{equation}\label{IIA1_PWaves}
w^\pm_{E}[r]\equiv \xi_E \exp[\pm i k_E r]
	\quad\text{with}\quad
	k_E\equiv\sqrt{2 M E/\hbar^2}
\end{equation}
describe the free propagation of carriers inside the leads and the
scattering amplitudes, $S^{\a\b+}_E$ and $S^{\a\b-}_E
=\bar{S}^{\b\a+}_E$, account for transition between the terminals 
$\b$ and $\a$\footnote{Throughout this article, bars indicate complex
conjugation.}.
The coordinate $r_\b\geq 0$ parameterizes the lead $\b$ in radial 
direction and the factor
\begin{equation}
\xi_E\equiv \sqrt{(d k_E/ dE)/2\pi} = \sqrt{M/2\pi k_E \hbar^2} 
\end{equation} 
has been introduced for normalization \cite{Baranger1989}. 

The outgoing and incoming states as defined by the conditions
\eqref{IIA1_SSE} and \eqref{IIA1_SBC} obey the orthogonality relations
\begin{equation}\label{IIA1_OrthR}
\braket{\vp^{\b\pm}_{E'}}{\vp^{\a\pm}_{E}}= \d_{\a\b}\d_{E-E'}
\end{equation}
and form two complete bases of the single-particle Hilbert space 
$\mathcal{H}$; for simplicity, we assume that no bound states exist
inside the scattering region throughout this article.

The scattering states are not normalizable and carry a finite 
probability current. 
Therefore, they cannot be interpreted in the same manner as bound 
states, whose wave function corresponds to the probability amplitude 
for finding a particle at a given position. 
Instead, we may regard the scattering states as a quantum mechanical
description of a homogeneous sequence of carriers that emerge from a
distant source and travel through the system one by one before being 
absorbed by a distant sink \cite{Ballentine1998}. 
This interpretation does not imply that the states 
$\ket{\vp^{\a\pm}_E}$ represent more than one particle; 
it rather entails that they describe a large number of identical and
independent scattering experiments \cite{Schiff1968}. 
In this picture, the square modulus of the scattering amplitude 
$S^{\a\b+}_E$ is the probability for a carrier with energy $E$ that 
is injected into the terminal $\a$ to leave the system through the
terminal $\b$. 
Analogously, the square modulus of $S^{\a\b-}_E$ is the probability 
for a carrier with energy $E$ that escapes through the terminal $\a$
to originate form the terminal $\b$. 

\subsection{Scattering Amplitudes}
\begin{figure}
\begin{center}
\includegraphics[scale=1.055]{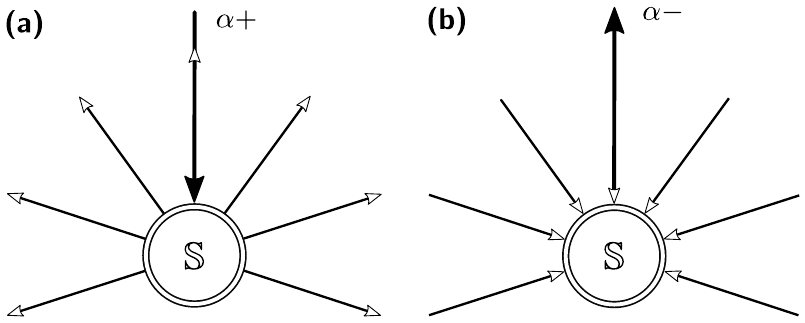}
\end{center}
\caption{Schematic representation of incoming and outgoing scattering
states. 
(a) The incoming state $\ket{\vp^{\a+}_E}$ consists of a plane wave 
with energy $E$ that approaches the sample in the terminal $\a$ and 
decomposes into $N$ escaping waves with the same energy but smaller 
amplitude. 
(b) The outgoing state $\ket{\vp^{\a-}_E}$ describes the time-reversed
situation, where $N$ approaching waves with the same energy $E$ 
combine into a single one that escapes through the terminal $\a$, cf. 
Eq.~\eqref{IIA1_SBC}. 
\label{Fig_2}}
\end{figure}
To ensure the conservation of probability currents, the scattering 
amplitudes have to obey the unitarity condition \cite{Baranger1989}
\begin{equation}\label{IIA2_UniC}
\gsum S^{\a\g\pm}_E S^{\g\b\mp}_E	= \d_{\a\b}. 
\end{equation}
Furthermore, they provide a link between outgoing and incoming states
by means of the relation
\begin{equation}\label{IIA2_InOutR}
\ket{\vp^{\a\pm}_E} = \bsum S^{\a\b\pm}_E\ket{\vp^{\b\mp}_E},
\end{equation}
which can be easily verified in position representation using the 
boundary conditions \eqref{IIA1_SBC} and Eq.~\eqref{IIA2_UniC}. 

Upon applying the orthogonality relation \eqref{IIA1_OrthR}, 
Eq.~\eqref{IIA2_InOutR} implies an algebraic expression for the 
scattering amplitudes in terms of the scattering states given by
\begin{equation}\label{IIA2_ScatAmplE}
\braket{\vp^{\b\mp}_{E'}}{\vp^{\a\pm}_{E}}
	= S^{\a\b\pm}_E\d_{E-E'}. 
\end{equation} 
This result makes it possible to establish a universal symmetry, 
which follows from the observation that outgoing and incoming states
are connected by time reversal, i.e., 
\begin{equation}\label{IIA2_ScatTRSym}
\ket{\vp^{\a\pm}_E}= \TR\ket{\tilde{\vp}^{\a\mp}_E}, 
\end{equation}
where $\TR$ denotes the anti-unitary time-reversal operator 
\cite{Mazenko2006} and tildes indicate the reversal of external
magnetic fields, see Fig.~\ref{Fig_2}. 
Consequently, we have 
\begin{equation}
\!\!\!\!\braket{\vp^{\b\mp}_{E'}}{\vp^{\a\pm}_{E}} 
	=\braket{\TR\tilde{\vp}^{\b\pm}_{E'}}{\TR\tilde{\vp}^{\a\mp}_{E}}
	=\braket{\tilde{\vp}^{\a\mp}_{E}}{\tilde{\vp}^{\b\pm}_{E'}} 
	=\tilde{S}^{\b\a\pm}_E \d_{E-E'}
\end{equation}
and therefore, given the Eqs.~\eqref{IIA2_ScatAmplE},
\begin{equation}\label{IIA2_AmplTRSym}
S^{\a\b\pm}_E = \tilde{S}^{\b\a\pm}_E. 
\end{equation}
Hence, for systems without magnetic fields, the scattering amplitudes 
for forward and backward transitions between any two terminals $\a$ and 
$\b$ are identical. 

\subsection{Lippmann-Schwinger Theory I: Autonomous Systems}
The scattering states and amplitudes can, in principle, be determined
by rewriting the stationary Schr\"odiner equation \eqref{IIA1_SSE} in
position representation, calculating the wave function inside the 
scattering region and matching it with the boundary conditions
\eqref{IIA1_SBC} \cite{Wagner1994,Al-Sahhar1999,Martinez2001}. 
This procedure, however, becomes impractical when the scattering 
wave functions cannot be found exactly and perturbation methods 
must be applied.
It is then more convenient to follow an algebraic approach, which we
develop next. 

We first divide the Hamiltonian \eqref{IIA1_Hamiltonian} into a free 
part $H_0$ and a perturbation $V$ acting only on the scattering 
region,  
\begin{equation}
H= H_0 + V, 
\end{equation}
where we assume that the scattering states for $H_0$ can be determined
exactly. 
Next, we combine the stationary Schr\"odinger equations for the free
and the perturbed scattering states, 
\begin{equation}\label{IIA3_SSE}
H_0\ket{\vp^{\a\pm}_{0E}} = E\ket{\vp^{\a\pm}_{0E}}
	\quad\text{and}\quad
	H\ket{\vp^{\a\pm}_E} = E\ket{\vp^{\a\pm}_E}
\end{equation}
into a single inhomogeneous linear equation,
\begin{equation}
[E-H_0][\ket{\vp^{\a\pm}_E}-\ket{\vp^{\a\pm}_{0E}}]
	= V\ket{\vp^{\a\pm}_E}. 
\end{equation}
This equation can be formally solved for the vector 
$\ket{\vp^{\a\pm}_E}-\ket{\vp^{\a\pm}_{0E}}$ after making the operator
$E-H_0$ invertible by adding a small imaginary shift. 
Following these steps, we arrive at the Lippmann-Schwinger equation 
\begin{equation}\label{IIA3_LSE}
\ket{\vp^{\a\pm}_E} = \ket{\vp^{\a\pm}_{0E}}
	+ [E-H_0\pm i\ve]^{-1}V\ket{\vp^{\a\pm}_E}, 
\end{equation}
where $\ve>0$ and the limit $\ve\rightarrow 0$ must be taken after
physical observables have been calculated. 
Note that the sign of the complex shift is important to ensure the
correct correspondence between free and perturbed outgoing and 
incoming states, for details see \cite{Ballentine1998}. 

By construction, the solutions $\ket{\vp^{\a\pm}_E}$ of 
Eq.~\eqref{IIA3_LSE} also solve the corresponding stationary 
Schr\"odinger equation.
However, the Lippmann-Schwinger equation contains more information,  
since it explicitly includes the continuity condition 
\begin{equation}
\lim\nolimits_{V\rightarrow 0}\ket{\vp^{\a\pm}_E}
	=\ket{\vp^{\a\pm}_{0E}},
\end{equation}
which ensures that the perturbed states $\ket{\vp^{\a\pm}_E}$ obey 
the boundary conditions \eqref{IIA1_SBC}.
That is, for a given set of free states $\ket{\vp^{\a\pm}_{0E}}$, the
Lippmann-Schwinger equation uniquely determines the outgoing and 
incoming states for the full Hamiltonian $H$, while the solutions of 
the stationary Schr\"odinger equation are unique only up to linear 
combinations of scattering states with the same energy 
\cite{Ballentine1998}. 

The Lippmann-Schwinger equation \eqref{IIA3_LSE} can be formally 
solved by iteration. 
This procedure yields 
\begin{align}
\label{IIA3_LSESol}
\ket{\vp^{\a\pm}_E} 
&=\sum\nolimits_{l=0}^\infty 
	\bigl[[E-H_0\pm i\ve]^{-1} V\bigr]^l\ket{\vp^{\a\pm}_{0E}}\\
&= \bigl[1-[E-H_0\pm i\ve]^{-1}V\bigr]^{-1}\ket{\vp^{\a\pm}_{0E}}
	\nonumber\\
&=\ket{\vp^{\a\pm}_{0E}} + [E- H\pm i\ve]^{-1}V\ket{\vp^{\a\pm}_{0E}}
	\nonumber
\end{align}
where the last line follows by noting that 
\begin{align}
& \bigl[1-[E-H_0\pm i\ve]^{-1}V\bigr]^{-1}\\
&=\bigl[ [E-H_0\pm i\ve]^{-1}[E-H\pm i\ve]\bigr]^{-1}\nonumber\\
&=[E-H\pm i\ve]^{-1}[E-H_0\pm i\ve]\nonumber\\
&= 1 + [E-H+\pm i\ve]^{-1}V.\nonumber
\end{align}
The expression \eqref{IIA3_LSESol} provides a systematic expansion of 
the scattering states $\ket{\vp^{\a\pm}_E}$ in terms of the 
perturbation $V$. 
Moreover, it implies that the solutions of the Lippmann-Schwinger 
equation obey the same orthogonality relation as the free states, 
since 
\begin{align}
\label{IIA3_OrthRPert}
\braket{\vp^{\b\pm}_{E'}}{\vp^{\a\pm}_{E}} &= 
	\braket{\vp^{\b\pm}_{0E'}}{\vp^{\a\pm}_{E}}
	+ \frac{\bra{\vp^{\b\pm}_{0E'}}V\ket{\vp^{\a\pm}_{E}}}{
	E'-E\mp i\ve}\\
	&=\braket{\vp^{\b\pm}_{0E'}}{\vp^{\a\pm}_{0E}}
	 =\d_{\a\b}\d_{E-E'}.\nonumber
\end{align}
Here, we have first inserted Eq.~\eqref{IIA3_LSESol} for 
$\ket{\vp^{\b\pm}_{E'}}$ and then Eq.~\eqref{IIA3_LSE} for
$\ket{\vp^{\a\pm}_{E}}$ in the first summand. 
Along the same lines, we find 
\begin{align}
\label{IIA3_ScatAmplEPert}
\braket{\vp^{\b\mp}_{E'}}{\vp^{\a\pm}_{E}} &=
\braket{\vp^{\b\mp}_{0E'}}{\vp^{\a\pm}_{E}}+\frac{\bra{\vp^{\b\mp}_{0E'}}V
	\ket{\vp^{\a\pm}_{E}}}{E'-E\pm i\ve}\\
&=\braket{\vp^{\b\mp}_{0E'}}{\vp^{\a\pm}_{0E}}\mp
	\frac{2i\ve}{(E-E')^2+\ve^2}\bra{\vp^{\b\mp}_{0E'}}V
	\ket{\vp^{\a\pm}_{E}}\nonumber\\
&=\bigl(S^{\a\b\pm}_{0E} \mp 2\pi i\bra{\vp^{\b\mp}_{0E'}}V
	\ket{\vp^{\a\pm}_{E}}\bigr)\d_{E-E'},\nonumber
\end{align}
where we have used the relation \cite{Appel2007}
\begin{equation}
\lim_{\ve\rightarrow 0}\frac{\ve}{a^2 + \ve^2}= 
	\pi\d_a,
\end{equation}
which must be understood in the sense of distributions, and
$S^{\a\b\pm}_{0E}$ denotes the scattering amplitudes for the free
Hamiltonian $H_0$.
Comparing Eq.~\eqref{IIA3_ScatAmplEPert} with Eq.~\eqref{IIA2_ScatAmplE}
yields the formula 
\begin{equation}
S^{\a\b\pm}_E = S^{\a\b\pm}_{0E}\mp 2\pi i\bra{\vp^{\b\mp}_{0E}}V
	\ket{\vp^{\a\pm}_{E}}, 
\end{equation}              
which makes it possible to calculate the full scattering amplitudes
order by order in $V$ using the expansion \eqref{IIA3_LSESol} of the 
scattering states $\ket{\vp^{\a\pm}_E}$. 
This perturbation scheme is a key result of the Lippmann-Schwinger 
formalism and will be developed further in the following section.  

\section{Floquet Scattering Theory}\label{Sec_FST_DC}
\subsection{Floquet Theory}
The carrier dynamics in a driven multi-terminal system is governed by
a Hamiltonian with the general form 
\begin{equation}\label{IIB1_H}
H_t = P^2/2M + U +V_t = H + V_t,
\end{equation}
where the dynamical potential $V_t$ accounts for time-dependent 
external fields acting on the scattering region.
If the driving is periodic with frequency $\omega\equiv2\pi/\tau$,
according to the Floquet theorem, the time dependent Schr\"odinger
equation admits a complete set of solutions that have the structure
\begin{equation}
\ket{\psi^\a_{E,t}} = \exp[-i E t/\hbar]\ket{\phi^\a_{E,t}},
\end{equation}
where $\ket{\phi^\a_{E,t+\tau}}= \ket{\phi^\a_{E,t}}$, the parameter
$E$  here plays the role of a continuous quantum number and $\a$ 
stands for any discrete quantum number \cite{Shirley1965,
ZelDovich1967,Sambe1973}.
The Floquet states $\ket{\phi^\a_{E,t}}$ obey the 
Floquet-Schr\"odinger equation
\begin{equation}\label{IIB1_FSE}
[H_t-i\hbar\partial_t]\ket{\phi^\a_{E,t}}= E\ket{\phi^\a_{E,t}}
\end{equation}
and form an orthogonal basis of the single-particle Hilbert space
$\mathcal{H}$ at every fixed time $t$.
 
In order to formulate a systematic scattering theory for periodically
driven systems, it is convenient to introduce the extended Hilbert 
space \cite{Sambe1973}
\begin{equation}
\hat{\mathcal{H}}\equiv\mathcal{H}\otimes\mathcal{H}_\tau, 
\end{equation}
where $\mathcal{H}_\tau$ denotes the Hilbert space of $\tau$-periodic
functions. 
In time representation, the elements $\Ket{\psi}$ of 
$\hat{\mathcal{H}}$ are $\tau$-periodic single-particle state vectors,
i.e., 
\begin{equation}
\langle t|\psi\rrangle = \ket{\psi_t}
	\quad\text{with}\quad \ket{\psi_{t+\tau}}=\ket{\psi_t}\in\mathcal{H}. 
\end{equation}
The scalar product in $\hat{\mathcal{H}}$ is defined as 
\begin{equation}\label{IIB1_SP}
\BraKet{\psi}{\chi} \equiv \frac{1}{\tau}\tint \braket{\psi_t}{\chi_t}. 
\end{equation}
This framework makes it possible to cast the Floquet-Schr\"odinger
equation \eqref{IIB1_FSE} into the form of a stationary Schr\"odinger
equation given by 
\begin{equation}\label{IIB1_SE}
\hat{H}\Ket{\phi^{m\a}_E} =E_m\Ket{\phi^{m\a}_E}
\quad\text{with}\quad E_m\equiv E+m\hbar\omega,
\end{equation}
where $m$ runs over all integers. 
The Floquet vectors $\Ket{\phi^{m\a}_E}$ are connected to the
Floquet states according to
\begin{equation}\label{IIB1_FV}
\langle t|\phi^{m\a}_E\rrangle = u^m_t\ket{\phi^{\a}_{E,t}}
\quad\text{with}\quad
u^m_t\equiv \exp[im\omega t]
\end{equation}
and the effective Hamiltonian $\hat{H}$, which is defined as 
\begin{equation}
\bra{t}\hat{H}\Ket{\psi} \equiv [H_t - i\hbar\partial_t]\ket{\psi_t}, 
\end{equation}
is a self-adjoint operator on $\hat{\mathcal{H}}$ with respect to the
scalar product \eqref{IIB1_SP}. 
The additional Fourier factor in Eq.~\eqref{IIB1_FV}, which is 
accounted for by the mode index $m$, was introduced to ensure 
that the solutions of Eq.~\eqref{IIB1_SE} are complete in 
$\hat{\mathcal{H}}$; 
this property will be required to develop an algebraic scattering 
theory in the extended Hilbert space. 
Once the Floquet vectors $\Ket{\phi^{m\a}_E}$ have been determined, a 
complete set of Floquet states $\ket{\phi^\a_{E,t}}$ that fulfill 
Eq.~\eqref{IIB1_FSE} is obtained by setting the mode index to zero and
returning to the time representation. 

\subsection{Lippmann-Schwinger Theory II: Driven Systems}
\label{Sec_IIB2}
Replacing the stationary Schr\"odinger equation \eqref{IIA1_SSE} with
Eq.~\eqref{IIB1_SE}, we can now extend the Lippmann-Schwinger 
theory of autonomous systems to systems with periodic driving. 
The dynamical potential $V_t$ thereby plays the role of the 
perturbation and the free states are replaced by the Floquet 
vectors
\begin{equation}\label{IIB2_FreeStates}
\langle t|\vp^{m\a\pm}_E\rrangle\equiv
	u^m_t\ket{\vp^{\a\pm}_E},
\end{equation}
where $\ket{\vp^{\a\pm}_E}$ are the scattering states for 
stationary part $H$ of the Hamiltonian \eqref{IIB1_H}. 
The free Floquet scattering vectors $\Ket{\vp^{m\a\pm}_E}$ form a
complete basis of the extended Hilbert space $\hat{\mathcal{H}}$,
for outgoing and incoming orientation respectively, and fulfill 
the Floquet-Schr\"odinger equation
\begin{equation}
\hat{H}_0\Ket{\vp^{m\a\pm}_E}= E_m\Ket{\vp^{m\a\pm}_E},
\end{equation}
where the free effective Hamiltonian is defined as 
\begin{equation}
\bra{t}\hat{H}_0\Ket{\psi} = [H-i\hbar\partial_t]\ket{\psi_t}. 
\end{equation}
Furthermore, using Eq.~\eqref{IIA1_OrthR} and Eq.~\eqref{IIB1_SP}, it
is straightforward to verify the orthogonality relation 
\begin{equation}
\BraKet{\vp^{n\b\pm}_{E'}}{\vp^{m\a\pm}_{E}}= \d_{mn}\d_{\a\b}\d_{E-E'}.
\end{equation}
Note that the quantum numbers $E$ and $\a$ have now been identified 
with the energy and the terminal of either an incident ($+$) or an 
escaping ($-$) carrier. 

The full Floquet scattering vectors $\Ket{\phi^{m\a\pm}_E}$ are those 
solutions of the Floquet-Schr\"odinger equation
\begin{equation}\label{IIB2_SE}
\hat{H}\Ket{\phi^{m\a\pm}_E} = E_m\Ket{\phi^{m\a\pm}_E}
\end{equation}
that reduce to the corresponding free vectors $\Ket{\vp^{m\a\pm}_E}$ 
in the stationary limit $V_t\rightarrow 0$. 
They are uniquely determined by the Floquet-Lippmann-Schwinger 
equation 
\begin{equation}\label{IIB2_FLSE}
\Ket{\phi^{m\a\pm}_E} = \Ket{\vp^{m\a\pm}_E} 
	+[E_m-\hat{H}_0\pm i\ve]^{-1}\hat{V}\Ket{\phi^{m\a\pm}_E},
\end{equation}
which can be derived along the same lines as Eq.~\eqref{IIA3_LSE}; 
the perturbation operator on the extended Hilbert space is thereby 
defined as $\langle t|\hat{V}\Ket{\psi} = V_t\ket{\psi_t}$. 
The formal solution of Eq.~\eqref{IIB2_FLSE} can be found by iteration
and reads 
\begin{align}
\label{IIB2_FLSESol}
\Ket{\phi^{m\a\pm}_E} & = \sum\nolimits_{k=0}^\infty\bigl[
	[E_m-\hat{H}_0\pm i\ve]^{-1}\hat{V}\bigr]^k\Ket{\vp^{m\a\pm}_E}\\
&= \bigl[1-[E_m-\hat{H}_0\pm i\ve]^{-1}\hat{V}\bigr]^{-1}
	\Ket{\vp^{m\a\pm}_E}\nonumber\\
&=\Ket{\vp^{m\a\pm}_E}+[E_m-\hat{H}\pm i\ve]^{-1}\hat{V}
	\Ket{\vp^{m\a\pm}_E}. \nonumber
\end{align}

Using the Eqs. \eqref{IIB2_FLSE} and \eqref{IIB2_FLSESol}, we can now 
establish the orthogonality relation for the Floquet scattering 
vectors,
\begin{align}
\label{IIB2_OrthRPert}
\BraKet{\phi^{n\b\pm}_{E'}}{\phi^{m\a\pm}_{E}} & =
	\BraKet{\vp^{n\b\pm}_{E'}}{\phi^{m\a\pm}_{E}}
	+\frac{\Bra{\vp^{n\b\pm}_{E'}}\hat{V}\Ket{\phi^{m\a\pm}_{E}}}{
	E'_n-E_m\mp i\ve}\\
&= \BraKet{\vp^{n\b\pm}_{E'}}{\vp^{m\a\pm}_{E}}
	=\d_{nm}\d_{\a\b}\d_{E-E'},\nonumber
\end{align}
and the connecting relations between outgoing and incoming vectors, 
\begin{align}
\label{IIB2_ScatAmplEPert}
& \BraKet{\phi^{n\b\mp}_{E'}}{\phi^{m\a\pm}_{E}} =
	\BraKet{\vp^{n\b\mp}_{E'}}{\phi^{m\a\pm}_{E}}
	+\frac{\Bra{\vp^{n\b\mp}_{E'}}\hat{V}\Ket{\phi^{m\a\pm}_{E}}}{
	E'_n-E_m \pm i\ve}\\
&= \BraKet{\vp^{n\b\mp}_{E'}}{\vp^{m\a\pm}_{E}}
	\mp\frac{2i\ve}{(E_m-E'_n)^2+\ve^2}
	\Bra{\vp^{n\b\mp}_{E'}}\hat{V}\Ket{\phi^{m\a\pm}_{E}}
	\nonumber\\
&= \bigl(\d_{mn} S^{\a\b\pm}_E 
	\mp 2\pi i\Bra{\vp^{n\b\mp}_{E\ix{m-n}}}\hat{V}\Ket{\phi^{m\a\pm}_{E}}\bigr)
	\d_{\Em-\Fn}.\nonumber
\end{align}
Here, we followed the same steps as in the derivations of the
Eqs.~\eqref{IIA3_OrthRPert} and \eqref{IIA3_ScatAmplEPert}. 
In the Eq.~\eqref{IIB2_ScatAmplEPert}, $S^{\a\b\pm}_E$ denotes the 
scattering amplitudes for the stationary Hamiltonian $H$.

\newcommand{\FS}{\mathcal{S}}
\subsection{Floquet Scattering Amplitudes I: General Properties}

The Floquet scattering amplitudes are defined as 
\begin{equation}\label{IIB3_FSADef}
\BraKet{\phi^{\phantom{\b}\!\!\!\!0\b\mp}_{E'}}{\phi^{m\a\pm}_{E}}
	  \equiv \FS^{\a\b\pm}_{m,E}\d_{\Em-E'}^{\phantom{\beta}},
\end{equation}
where $\FS^{\a\b-}_{m,E}=\bar{\FS}^{\b\a+}_{-m,\Em}$.
They satisfy the unitarity conditions
\begin{align}
\label{IIB3_UniC}
\sum\nolimits_{n\g}\FS^{\g\b\mp}_{m-n,\En}
	\FS^{\a\g\pm}_{n,E}=\d_{m0}\d_{\a\b}
\end{align}
and the symmetry relation
\begin{equation}\label{IIB3_FSATRSym}
\FS^{\a\b\pm}_{m,E} = \dbtilde{\FS}^{\b\a\pm}_{-m,\Em},
\end{equation}
where the double tilde indicates the reversal of both external 
magnetic fields and driving protocols. 
In the following, we will show how these results can be derived within
the framework of Floquet scattering theory. 
Note that, throughout this article, we understand that sums over the
mode index run over all integers and that the Floquet scattering 
amplitudes are zero if their energy argument is not positive. 

The unitarity conditions \eqref{IIB3_UniC} follow from the 
completeness relation for the Floquet scattering vectors,
\begin{equation}\label{IIB3_CompR}
\sint{m\a}{E}\Ket{\phi^{m\a\pm}_E}\Bra{\phi^{m\a\pm}_E} = 1,
\end{equation}
where $1$ stands for the identity operator on the extended Hilbert 
space and the symbolic notation
\begin{equation}
\sint{m\a}{E}\equiv\sum\nolimits_{m\a}\Eint
\end{equation}
has been introduced for convenience.  
We thus have\footnote{Note that $\BraKet{\phi^{n\g\mp}_{E''}}{
\phi^{m\a\pm}_E}=\BraKet{\phi^{0\g\mp}_{E''}}{\phi^{m-n\a\pm}_{E}}$.}
\begin{align}
\BraKet{\phi^{0\b\pm}_{E'}}{\phi^{m\a\pm}_E} 
	&= \d_{m0}\d_{\a\b}\d_{\Em-E'}\\
&=\sint{n\g}{E''} \BraKet{\phi^{0\b\pm}_{E'}}{\phi^{n\g\mp}_{E''}}
	\BraKet{\phi^{n\g\mp}_{E''}}{\phi^{m\a\pm}_E}\nonumber\\
&=\Bigl(\sum\nolimits_{n\g} \FS^{\g\b\mp}_{n,E\ix{m-n}}
	\FS^{\a\g\pm}_{m-n,E}\Bigr)\d_{\Em-E'}.\nonumber
\end{align}
and shifting the summation index $p$ yields the result
\eqref{IIB3_UniC}. 

To derive the symmetry relation \eqref{IIB3_FSATRSym}, we first 
observe that the free outgoing and incoming Floquet scattering vectors
are connected by time reversal, i.e.,
\begin{equation}
\hat{\TR}\Ket{\vp^{m\a\pm}_E} = \Ket{\tilde{\vp}^{m\a\mp}_E}, 
\end{equation}
as can be easily verified with the help of Eq.~\eqref{IIB2_FreeStates}
and the definition of the time-reversal operator on the extended 
Hilbert space, $\bra{t}\hat{\TR}\Ket{\psi}\equiv\TR\ket{\psi_{-t}}$. 
Consequently, acting on the solution of the Floquet-Lippmann-Schwinger
equation, Eq.~\eqref{IIB2_FLSESol}, with $\hat{\TR}$ yields\footnote{
Recall that a single tilde indicates the reversal of magnetic
fields only and a double tilde includes the reversal of driving 
protocols.}
\begin{equation}\label{IIB3_ScatVecTRSym}
\hat{\TR}\Ket{\phi^{m\a\pm}_E}= \Ket{\dbtilde{\phi}^{m\a\mp}_E},
\end{equation}
where we have used the identity $\hat{\TR}\hat{V} =\dbtilde{V}
\hat{\TR}$ with the time-reversed perturbation operator being 
defined as $\bra{t}\dbtilde{V}\Ket{\psi}\equiv \tilde{V}_{-t}
\ket{\psi_t}$. 
This result finally implies 
\begin{align}
\BraKet{\phi^{0\b\mp}_{E'}}{\phi^{m\a\pm}_{E}}
	&=\BraKet{\hat{\TR}\dbtilde{\phi}^{0\b\pm}_{E'}}{
	\hat{\TR}\dbtilde{\phi}^{m\a\mp}_{E}}\\
	&=\BraKet{\dbtilde{\phi}^{m\a\mp}_{E}}{
	\dbtilde{\phi}^{0\b\pm}_{E'}}
	=\dbtilde{S}^{\b\a\pm}_{-m,E\ix{m}}\d_{\Em-E'}
	\nonumber
\end{align}
and thus, by comparison with the definition \eqref{IIB3_FSADef},
the symmetries \eqref{IIB3_FSATRSym}. 

\subsection{Floquet Scattering Amplitudes II: Perturbation Theory}\label{Sec_PertTh}

The framework of our Floquet-Lippmann-Schwinger theory makes it 
possible to derive a systematic expansion of the Floquet scattering
amplitudes in powers of the dynamical potential.
To this end, we first compare the definitions \eqref{IIB3_FSADef} with
the relations~\eqref{IIB2_ScatAmplEPert} to obtain the explicit 
expressions 
\begin{equation}\label{IIB3_FSAExpl}
\FS^{\a\b\pm}_{m,E} = \d_{m0} S^{\a\b\pm}_{E} 
	\mp 2\pi i \Bra{\vp^{0\b\mp}_{\Em}}\hat{V}
	\Ket{\phi^{m\a\pm}_{E}}.
\end{equation}
Inserting the series representation \eqref{IIB2_FLSESol} of the 
Floquet scattering vector $\Ket{\phi^{m\a\pm}_E}$ into this
formula now yields the expansion
\begin{equation}
\begin{multlined}[t][\displaywidth]
S^{\a\b\pm}_{m,E} =  \d_{m0} S^{\a\b\pm}_E\\
\mp 2\pi i \sum\nolimits_{l=0}^\infty
	\Bra{\vp^{0\b\mp}_{\Em}}\hat{V}
	\bigl[[\Em-\hat{H}_0\pm i\ve]^{-1}\hat{V}\bigr]^l
	\Ket{\vp^{m\a\pm}_E},
\end{multlined}
\end{equation}
This result is analogous to the Born series in standard 
scattering theory \cite{Newton1982}. 
Taking into account only first-order corrections gives the 
Floquet-Born approximation 
\begin{align}
S^{\a\b\pm}_{m,E} &\simeq 
	\d_{m0} S^{\a\b\pm}_E \mp 2\pi i \Bra{\vp^{0\b\mp}_{\Em}}
	\hat{V}\Ket{\vp^{m\a\pm}_E}\\
&= \d_{m0} S^{\a\b\pm}_{E} \mp\frac{2\pi i}{\tau}\tint
	\bra{\vp^{\b\mp}_{\Em}}V_t\ket{\vp^{\a\pm}_{E}}u^{m}_t,
	\nonumber
\end{align}
which is justified if the amplitude of the external potential 
variations are small compared to the carrier energy. 

\subsection{Scattering Wave Functions}\label{Sec_IIB4}
The physical content of the Floquet scattering states can be
understood from their asymptotic wave functions. 
To derive their structure, we first use the Floquet-Lippmann-Schwinger
\eqref{IIB2_FLSE} and the completeness relation for the free Floquet 
scattering vectors, 
\begin{equation}
\sint{m\a}{E} \Ket{\vp^{m\a\pm}_E}\Bra{\vp^{m\a\pm}_E} = 1, 
\end{equation}
to connect the lead wave functions of the Floquet scattering states 
with the lead wave functions \eqref{IIA1_SBC} of the stationary 
scattering states, 
\begin{align}
\label{IIB4_SBCAux1}
&\phi^{\a\pm}_{E,t}[r_\b]\equiv \braket{r_\b}{\phi^{\a\pm}_{E,t}}
	=\BraKet{r_\b,t}{\phi^{0\a\pm}_E}\\
&=\BraKet{r_\b,t}{\vp^{0\a\pm}_E}
	+\Bra{r_\b,t}[E-\hat{H}_0\pm i\ve]^{-1}\hat{V}\Ket{\phi^{0\a\pm}_E}
	\nonumber\\
&=\vp^{\a\pm}_E[r_\b]
	+\sint{m\g}{E'} \vp^{\g\mp}_{E'}[r_\b]u^m_t
	\frac{\Bra{\vp^{m\g\mp}_{E'}}\hat{V}\Ket{\phi^{0\a\pm}_E}}{
	E-E'_m\pm i\ve}.\nonumber
\end{align}
This expression shows that the wave functions
$\phi^{\a\pm}_{E,t}[r_\b]$ are invariant under spatial translations by
integer multiples of the wave length $\lambda_E \equiv 2\pi/k_E$. 
Therefore, we can evaluate them in the far distance from the 
scattering region.
Plugging Eq.~\eqref{IIA1_SBC} into \eqref{IIB4_SBCAux1} thus yields 
\begin{equation}
\begin{multlined}[t][\displaywidth]
\phi^{\a\pm}_{E,t}[r_\b] = \d_{\a\b} 
	w^\mp_E[r_\b]+S^{\a\b\pm}_E w^\pm_E[r_\b]\\
\shoveright{-\sint{m\g}{E'} 
	\frac{\d_{\b\g}w^\pm_{E'}[r_\b]+S^{\g\b\mp}_{E'}w^\mp_{E'}[r_\b]}{
	E'-E_m \mp i\ve}}\\
\shoveright{\times u^{-m}_t\Bra{\vp^{-m\g\mp}_{E'}}\hat{V}
	\Ket{\phi^{0\a\pm}_E}}\\[6pt]
\shoveleft{\asymp \d_{\a\b} w^\mp_E[r_\b]
	+\msum \bigl(\d_{m0} S^{\a\b\pm}_E\mp 2\pi i \Bra{\vp^{0\a\mp}_{\Em}}
	\hat{V}\Ket{\phi^{m\a\pm}_E}\bigr)}\\
\times w^\pm_{\Em}[r_\b] u^{-m}_t
\end{multlined}
\end{equation}
Here, we have used \emph{Lemma~1c} of App.~\ref{ApxA} and the symbol
$\asymp$ indicates asymptotic equality in the limit $r_\a\rightarrow
\infty$.  
Finally, inserting the expressions \eqref{IIB3_FSAExpl} for the 
Floquet scattering amplitudes gives the wave function
\begin{equation}\label{IIB4_SBC}
\!\!\!\phi^{\a\pm}_{E,t}[r_\b] = 
	\d_{\a\b}w^\mp_E[r_\b]
	+\msum \FS^{\a\b\pm}_{m,E} w^\pm_{\Em}[r_\b]u^{-m}_t.
\end{equation}

This result shows that the outgoing and incoming Floquet scattering 
states, $\ket{\phi^{\a+}_{E,t}}$ and $\ket{\phi^{\a-}_{E,t}}$,
respectively, contain a single incident and escaping wave with wave 
length $\lambda_E$ in the lead $\a$.
Hence, they represent a carrier with energy $E$ that either enters or
leaves the system through the terminal $\a$. 
The Floquet scattering amplitude $S^{\a\b+}_{m,E}$ thus corresponds to
the probability amplitude for a transitions from the terminal $\a$
to the terminal $\b$ under the absorption $(m>0)$ or emission $(m<0)$
of $m$ units of energy $\hbar\omega$. 
Analogously, $S^{\a\b-}_{m,E}$ corresponds to the probability 
amplitude for that an escaping carrier with energy $E$ in the terminal
$\a$ was injected into the terminal $\b$ with an energy surplus 
$(m>0)$ or deficit $(m<0)$ of $m$ quanta $\hbar\omega$.  
In this picture, the unitarity condition \eqref{IIB3_UniC} ensures the
conservation of probability currents. 
The symmetry relation \eqref{IIB3_FSATRSym} implies that forward and 
backward processes occur with the same probability amplitude 
provided that no magnetic field is applied to the system and the 
driving protocols are invariant under time reversal
\cite{Moskalets2002,Moskalets2004}. 

We stress that the lead wave functions \eqref{IIB4_SBC} have not been
used to define the Floquet scattering states in our approach;
in fact, their structure results from the continuity condition 
$\lim\nolimits_{V\!\ix{t}\rightarrow 0}
\ket{\phi^{\a\pm}_{E,t}}=\ket{\vp^{\a\pm}_E}$, which has been built 
into the Floquet-Lippmann-Schwinger equation \eqref{IIB2_FLSE}. 
In the same way, the quantization of the energy flux between carriers
and driving fields arises naturally from the periodicity condition 
$\ket{\phi^{\a\pm}_{E,t}}=\ket{\phi^{\a\pm}_{E,t+\tau}}$,
which is imposed by the Floquet theorem and encoded in structure
of the extended Hilbert space.

\section{Matter and Energy Currents}\label{Sec_MaEnCurr}
\subsection{Current Operators}
On the single-particle level, the matter and energy currents that 
flow at the position $r_\a$ of the lead $\a$ into a multi-terminal
conductor are represented by the operators \cite{Hardy1963,Kugler1967}
\begin{subequations}\label{IIIA_CurrOp1}
\begin{align}
\label{IIIA_PaCurrOp1}
j_\a^{\rho} &\equiv -\frac{1}{2M}\{P,\d[R-r_\a]\}
	\quad\text{and}\\[6pt]
\label{IIIA_EnCurrOp1}
j_\a^{\E} &\equiv-\frac{1}{8M} \bigl\{P^2,\{P,\d[R-r_\a]\}
	\bigr\}.
\end{align}
\end{subequations}
Here, $R$ and $P$ are the position and momentum operators, $M$ 
denotes the carrier mass and curly brackets indicate the usual 
anti-commutator. 
Note that, for convenience, we notationally suppress the dependence
of the current operators on the coordinate $r_\a$ throughout. 

Since the transport carriers are indistinguishable, the many-body
quantum state of a mesoscopic conductor must be either symmetric or 
antisymmetric under the exchange of two arbitrary carriers. 
An elegant method to take this constraint into account is provided by 
the language of second quantization, which can be adopted to our
present setup as follows. 
We first introduce the scattering field operators $\sPhi^{\a}_{E,t}$
and $\sPhi^{\a\dagger}_{E,t}$, which annihilate and create a carrier
in the outgoing Floquet scattering state $\ket{\phi^{\a+}_{E,t}}$, 
respectively. 
For any fixed time $t$, these operators obey the commutation relations
\begin{subequations}\label{IIIA_ComRules}
\begin{align}
&\bigl\{\sPhi^{\b}_{E',t},\sPhi^{\a}_{E,t}\bigr\}
	=\bigl\{\sPhi^{\b\dagger}_{E',t},\sPhi^{\a\dagger}_{E,t}\bigr\}=0
	\quad\text{and}\\[6pt]
&\bigl\{\sPhi^{\b\dagger}_{E',t},\sPhi^{\a}_{E,t}\bigr\}
	=\bigl\{\sPhi^{\b}_{E',t},\sPhi^{\a\dagger}_{E,t}\bigr\}
	= \d_{\a\b}\d_{E-E'},
\end{align}
\end{subequations}
where we focus on Fermions for the sake of concreteness; 
the theory for Bosonic carriers can be developed analogously.
The many-particle current operators can now be expresses as 
\begin{equation}\label{IIIA_CurrOpSP}
\Jf^x_\a = \sint{\;\b}{E}\sint{\;\g}{E'}
	j^{x\a,\b\g}_{EE',t}\sPhi^{\b\dagger}_{E,t}\sPhi^{\g}_{E',t},
\end{equation} 
where $x\equiv\rho,\E$ and
\begin{subequations}
\begin{align}
\label{IIIA_PaCurrME}
&j^{\rho\a,\b\g}_{EE',t}\equiv\bra{\phi^{\b+}_{E,t}}\;j^{\rho}_\a
	\ket{\phi^{\g+}_{E',t}}\\
&\phantom{j^{\rho\a,\b\g}_{EE',t}}=\frac{i\hbar}{2m}\left.\left(
	\bar{\phi}^{\b}_{E}	\phi^{\g}_{E';1}
	-\phi^{\g}_{E'}\bar{\phi}^{\b}_{E;1}
	\right)\right|_{r=r\ix{\a}}\nonumber\\
\label{IIIA_EnCurrME}
&j^{\E\a,\b\g}_{EE',t}\equiv\bra{\phi^{\b+}_{E,t}}\;j^{\E}_\a
	\ket{\phi^{\g}_{E',t}}\\
&=\frac{i\hbar^3}{8m^2}\left.\left(
	\phi^{\g}_{E'}\bar{\phi}^{\b}_{E;3}
	-\phi^{\g}_{E';1}	\bar{\phi}^\b_{E;2}
	+\bar{\phi}^{\b}_{E;1}\phi^{\g}_{E';2}
	-\bar{\phi}^{\b}_{E}\phi^{\g}_{E';3}\right)\right|_{r=r\ix{\a}}
	\nonumber
\end{align}
\end{subequations}
with $\phi^{\a}_{E} \equiv\phi^{\a+}_{E,t}[r]$ and $\phi^{\a}_{E;l}
\equiv\partial_r^l\phi^{\a+}_{E,t}[r]$. 
These matrix elements are $\tau$-periodic functions of $t$ and can
thus be expanded in a Fourier series, 
\begin{equation}\label{IIIA_CurrMEFourier}
j^{x\a,\b\g}_{EE',t} \equiv \msum \jc^{x\a,\b\g}_{EE',m}
	\exp[im\omega t], 
\end{equation}
where the coefficients $\jc^{x\a,\b\g}_{EE',m}$ can be determined from
the Floquet scattering wave functions \eqref{IIB4_SBC}.
Rather than spelling out the corresponding expressions in full 
generality, we here provide only a specific set of Fourier components
that will be needed in the following sections and can be written in 
the compact form
\begin{align}\label{IIIA_CurrMEEx}
&\jc^{x\a,\b\g}_{E\Em,m}= \frac{1}{h}\Bigl(
	\d_{\a\b}\d_{\a\g}\d_{m0}\zeta_E^x
  	-\nsum \FS^{\g\a+}_{n-m,\Em}\bar{\FS}^{\b\a+}_{n,E}
  	\zeta^x_{\En}\Bigr)\nonumber\\[6pt]
& \text{with}\quad h\equiv 2\pi\hbar,\quad
	\zeta^\rho_E\equiv 1,\quad\zeta^\E_E\equiv E. 
\end{align}

\subsection{Mean Currents}
We are now ready to calculate the average steady-state currents of
matter and energy in a periodically driven multi-terminal conductor. 
To this end, we recall the general formula \eqref{Int_CurrCorr} for
the mean currents,
\begin{equation}\label{IIIB_MeanCurrGen}
J^x_\a \equiv\limt\bigl\langle  \Jf^x_{\a,t'}\bigr\rangle. 
\end{equation} 
The Heisenberg-picture operator $\Jf^x_{\a,t}$ thereby describes the 
flow of particles $(x=\rho)$ or energy $(x=\E)$ at a given time $t$ 
and at a given position $r_\a$ in the lead $\a$; angular brackets 
denote the ensemble average over all possible quantum states of the 
system. 

The formula \eqref{IIIB_MeanCurrGen} can be evaluated in two steps.
First, transforming the current operators \eqref{IIIA_CurrOpSP} into 
the Heisenberg picture yields 
\begin{align}
\label{IIIB_CurrOpHP}
\Jf^x_{\a,t} &\equiv \Uf_{t}^\dagger \Jf^x_\a \Uf_t\\
& = \sint{\;\b}{E}\sint{\;\g}{E'} 
	j^{x\a,\b\g}_{EE',t} \sPhi^{\b\dagger}_{E}\sPhi^{\g}_{E'}
	\exp[i(E-E')t/\hbar],\nonumber
\end{align} 
where the unitary operator $\Uf_{t}$ generates the evolution of the 
many-particle system from the time $0$ to the time $t$. 
The second line in Eq.~\eqref{IIIB_CurrOpHP} follows from the time 
evolution laws for the field operators, 
\begin{subequations}
\begin{align}
\label{IIIB_EvLawFOa}
&\Uf^\dagger_t \sPhi^{\a}_{E,t}\Uf_t 
	=\sPhi^\a_E\exp[-i Et/\hbar]
	\quad\text{and}\\[6pt]
\label{IIIB_EvLawFOb}
&\Uf^\dagger_t \sPhi^{\a\dagger}_{E,t}\Uf_t
	=\sPhi^\a_E\exp[i Et/\hbar], 
\end{align}
\end{subequations}
which, in turn, are a consequence of the fact that the outgoing
scattering states $\ket{\phi^{\a+}_{E,t}}$ are solutions of the 
Floquet Schr\"odinger equation \eqref{IIB1_FSE} and thus 
fulfill\footnote{To verify the time evolution laws for the scattering
field operators, construct a basis of the many-particle Fock space
from the incoming Floquet scattering states $\ket{\phi^{\a+}_E}$
and evaluate the corresponding matrix elements of both sides of the 
Eqs.~\eqref{IIIB_EvLawFOa} and \eqref{IIIB_EvLawFOb} with the help of
the relation~\eqref{IIIB_EvLawFSS}.}
\begin{equation}\label{IIIB_EvLawFSS}
U_t\ket{\phi^{\a+}_{E}} 
	= \exp[-iEt/\hbar]\ket{\phi^{\a+}_{E,t}}. 
\end{equation}
Here, $U_t$ is the single-particle time evolution operator. 
Note that the time argument $0$ is omitted throughout for simplicity.

Second, to evaluate the ensemble average in 
Eq.~\eqref{IIIB_MeanCurrGen}, we recall that the outgoing Floquet 
scattering states $\ket{\phi^{\a+}_{E,t}}$ are populated with 
non-interacting carriers by a thermochemical reservoir with 
temperature $T_\a$ and chemical potential $\mu_\a$. 
Hence, provided that all reservoirs are  mutually independent, the 
quantum-statistical average of an ordered pair of one creation and one
anihilation operator is given by the grand canonical rule
\begin{align}\label{IIIB_ThAv}
&\bigl\langle\sPhi^{\a\dagger}_E \sPhi^\b_{E'}\bigr\rangle 
	= \d_{\a\b}\d_{E-E'}f^\a_E,\\[6pt]
&\text{where}\quad f^\a_E\equiv \frac{1}{1+\exp[(E-\mu_\a)/T_\a]}
\nonumber
\end{align}
denotes the Fermi function of the reservoir $\a$ and Boltzmann's
constant is set to $1$ throughout; 
averages of products that contain different numbers of creation and 
annihilation are zero \cite{Buttiker1992,Blanter2000,Moskalets2004}. 

Inserting Eq.~\eqref{IIIB_CurrOpHP} into the formula 
\eqref{IIIB_MeanCurrGen} and using Eq.~\eqref{IIIB_ThAv} yields
\begin{equation}\label{IIIB_MCAux}
\!\! J^x_\a = \limt\!\!\sint{\b}{E} j^{x\a,\b\b}_{EE,t} f^\b_E
	=\sint{\b}{E} \jc^{x\a,\b\b}_{EE,0} f^\b_E,
\end{equation}
where we have used the Fourier expansion 
\eqref{IIIA_CurrMEFourier} for the second identity. 
Upon recalling the matrix elements \eqref{IIIA_CurrMEEx}, the mean 
currents can now be expressed in terms of the Floquet scattering 
amplitudes of the conductor and the Fermi functions of the attached 
reservoirs, 
\begin{equation}\label{IIIB_MC}
J^x_\a = \frac{1}{h}\sint{\b}{E} 
	\bigl(\d_{\a\b}\zeta^x_E -\msum |\FS^{\b\a+}_{m,E}|^2 \zeta^x_{\Em}
	\bigr) f^\b_E.  
\end{equation}
This formula, which holds arbitrary far from equilibrium, shows that
the conductance properties of a coherent multi-terminal system are 
fully determined by its Floquet scattering amplitudes. 
In the limit $V_t\rightarrow 0$, where the Floquet scattering 
amplitudes become equal to the stationary ones according to 
Eq.~\eqref{IIB3_FSAExpl}, it reduces to the standard
Landauer-B\"uttiker formula. 

The physical consistency of the current formula \eqref{IIIB_MC} 
derives from the sum rules 
\begin{equation}\label{IIIB_SumRules}
\sum\nolimits_{m\a}|\FS^{\a\b+}_{-m,\Em}|^2=1 \quad\text{and}\quad
	\sum\nolimits_{m\a}|\FS^{\b\a+}_{m,E}|^2=1, 
\end{equation}
which follow directly from the unitarity conditions for the Floquet
scattering amplitudes, Eq.~\eqref{IIB3_UniC}.
By using the first of these relations, Eq.~\eqref{IIIB_MC} can be 
rewritten in the form
\begin{equation}\label{IIIB_MCDiff}
J^x_\a = \frac{1}{h}\sint{m\b}{E}\;|\FS^{\b\a+}_{m,E}|^2\zeta^x_{\Em}
	(f^\a_{\Em}-f^\b_E). 
\end{equation}
This result shows that the mean currents indeed vanish in equilibrium,
i.e., if all reservoirs are at the same temperature and chemical 
potential and the external driving fields are turned off. 
Furthermore, by summing both sides of Eq.~\eqref{IIIB_MC} over the 
terminal index and using the second sum rule in 
Eq.~\eqref{IIIB_SumRules}, we recover the fundamental conservation 
laws for matter and energy, 
\begin{equation}\label{IIIB_ConLaws}
\asum J^\rho_\a = 0
	\quad\text{and}\quad
	\asum J^\E_\a = -\Pi_{{{\rm ac}}}.
\end{equation}
The average power that is injected into the system through the 
external driving, $\Pi_{{{\rm ac}}}$, thereby admits the 
microscopic expression
\begin{equation}
\Pi_{{{\rm ac}}}\equiv 
	\frac{1}{\tau}\sint{m\a\b}{E}\;|\FS^{\b\a+}_{m,E}|^2 
	m f^\b_E. 
\end{equation}

\subsection{Zero-Frequency Noise}\label{Sec_CN}
The zero-frequency noise, or noise power, of the matter and energy
currents in a multi-terminal conductor is given by the general formula
\begin{align}\label{IIIC_CurrCorr}
P^{xy}_{\a\b}
&\equiv\limtt\bigl\langle
	(\Jf^x_{\a,t'}-J^x_\a)(\Jf^y_{\b,t''}-J^y_\b)\bigr\rangle\\
&=\limtt\bigl\llangle \Jf^x_{\a,t'};\Jf^y_{\b,t''}\bigr\rrangle
	\nonumber
\end{align}
for $x=\rho,\E$ and $y=\rho,\E$. 
Here, the notation
\begin{equation}\label{IIIC_CorrF}
	\llangle A;B\rrangle
	\equiv \langle A B\rangle
	-\langle A\rangle\langle B\rangle
\end{equation}
has been introduced for the correlation function of the observables 
$A$ and $B$. 
The quantity $P^{xy}_{\a\b}$ can be calculated with the same
techniques as the mean currents.  
In the first step, we use Eq.~\eqref{IIIB_CurrOpHP} to express the 
time-dependent current operators in terms of the scattering field 
operators and obtain
\begin{equation}\label{IIIC_CCAux1}
\begin{multlined}[t][\displaywidth]
P^{xy}_{\a\b}=
	\limtt\!\!
	\sint{\g\ix{1}}{E\ix{1}}\cdots\!\sint{\g\ix{4}}{E\ix{4}}\\[3pt]
\shoveright{
	j^{x\a,\g\ix{1}\g\ix{2}}_{E\ix{1}E\ix{2},t'}\cdot
	j^{y\b,\g\ix{3}\g\ix{4}}_{E\ix{3}E\ix{4},t''}
	\bigl\llangle\sPhi^{\g\ix{1}\dagger}_{E\ix{1}}
	\sPhi^{\g\ix{2}}_{E\ix{2}};
	\sPhi^{\g\ix{3}\dagger}_{E\ix{3}}
	\sPhi^{\g\ix{4}}_{E\ix{4}}\bigr\rrangle}\\[3pt]
	\times\exp[i(E_1-E_2)t'/\hbar]\exp[i(E_3-E_4)t''/\hbar]
	\end{multlined}
\end{equation}
with dots being inserted to improve readability. 
The correlation function of the scattering field operators
in Eq.~\eqref{IIIC_CCAux1} can be evaluated using the 
finite-temperature version of Wick's theorem \cite{Giulianni2005},
which implies 
\begin{align}\label{IIIC_Wick}
&\bigl\llangle
	\sPhi^{\g\ix{1}\dagger}_{E\ix{1}}
	\sPhi^{\g\ix{2}}_{E\ix{2}};
	\sPhi^{\g\ix{3}\dagger}_{E\ix{3}}
	\sPhi^{\g\ix{4}}_{E\ix{4}}\bigr\rrangle
	=\bigl\langle\sPhi^{\g\ix{1}\dagger}_{E\ix{1}}
	\sPhi^{\g\ix{4}}_{E\ix{4}}
	\bigr\rangle\bigl\langle	\sPhi^{\g\ix{2}}_{E\ix{2}}
	\sPhi^{\g\ix{3}\dagger}_{E\ix{3}}\bigr\rangle\\[3pt]
&=\d_{\g\ix{1}\g\ix{4}}\d_{E\ix{1}-E\ix{4}}f^\g_{E\ix{1}}
	\cdot\d_{\g\ix{2}\g\ix{3}}\d_{E\ix{2}-E\ix{3}}(1-f^\d_{E\ix{2}}).
	\nonumber
\end{align}
Here, we have used the commutation rules \eqref{IIIA_ComRules} and 
the grand canonical averaging rule \eqref{IIIB_ThAv} for the last 
identity. 
After inserting Eq.~\eqref{IIIC_Wick} and the Fourier expansion of the 
current matrix elements \eqref{IIIA_CurrMEFourier} into
Eq.~\eqref{IIIC_CCAux1}, we can carry out the time integrals. 
This step yields
\begin{equation}\label{IIIC_CCAux2}
\begin{multlined}[t][.8\displaywidth]
P^{xy}_{\a\b} = \lim_{t\rightarrow\infty} 
	\sint{m\g}{E}\sint{n\d}{E'}\jc^{x\a,\g\d}_{EE',m}\cdot
	\jc^{y\b,\d\g}_{E'E,-n}\cdot f^\g_E (1-f^\d_{E'})\\[3pt]
\times t\cdot 
	\frac{\exp[-i(E'-E_m)t/\hbar]-1}{(E'-E_m)t/\hbar}
	\frac{\exp[i(E'-E_n)t/\hbar]-1}{(E'-E_n)t/\hbar}. 
\end{multlined}
\end{equation}
Upon taking the limit $t\rightarrow\infty$ with the help of 
\emph{Lemma~2} of App.~\ref{ApxA}, this expression simplifies to the
compact result
\begin{equation}\label{IIIC_CCAux3}
P^{xy}_{\a\b} = h \sint{m\g\d}{E}\;
	\jc^{x\a,\g\d}_{E\Em,m}\cdot\bar{\jc}^{y\b,\g\d}_{E\Em,m}\cdot 
	f^\g_E (1-f^\d_{\Em}),
\end{equation} 
where we have applied the relation $\jc^{y\b,\d\g}_{\Em E,-m} 
=\bar{\jc}^{y\b,\g\d}_{E\Em,m}$.

The zero-frequency noise can now be expressed in terms of the Floquet
scattering amplitudes of the driven conductor and the 
Fermi functions of the reservoirs. 
To this end, we insert the matrix elements \eqref{IIIA_CurrMEEx} into
Eq.~\eqref{IIIC_CCAux3}. 
After some algebra, we thus obtain the explicit formula
\begin{equation}\label{IIIC_CurrCorrEx}
\begin{multlined}[t][\displaywidth]
P^{xy}_{\a\b} = \frac{1}{h}\sint{m}{E}\Bigl(
	\d_{m0}\d_{\a\b}\zeta^x_E\zeta^y_E f'^\a_E\\[3pt]
\shoveright{-|\FS^{\a\b+}_{m,E}|^2\zeta^x_E\zeta^y_{\Em} f'^\a_E
	-|\FS^{\b\a+}_{m,E}|^2\zeta^x_{\Em}\zeta^y_E f'^\b_E}\\[3pt]
	+\sum\nolimits_{\g\d}
	\Ac^{x\a,\g\d}_{m,E}\bar{\Ac}^{y\b,\g\d}_{m,E}
	f^\g_E(1-f^\d_{\Em})\Bigr),
\end{multlined}
\end{equation}
where we have introduced the abbreviations 
\begin{equation}
f'^\a_E\equiv f^\a_E(1-f^\a_E),\;\;\;
	\Ac^{x\a,\g\d}_{m,E}\equiv 
	\nsum \bar{\FS}^{\g\a+}_{n,E}\FS^{\d\a+}_{n-m,\Em}\zeta^x_{\Em}
\end{equation}
for convenience.\footnote{
The formula \eqref{IIIC_CurrCorrEx} shows that the noise power 
$P^{xy}_{\a\b}$ is real and obeys the symmetry $P^{xy}_{\a\b} = 
P^{yx}_{\b\a}$. 
These properties cannot be \emph{a priori} expected as the current
correlation function in Eq.~\eqref{IIIC_CurrCorr} is, in general,
not symmetric with respect to the current operators.
In fact, the antisymmetric, imaginary part of this correlation
function is wiped out only when the limit $t\rightarrow\infty$ is 
taken in Eq.~\eqref{IIIC_CCAux2}. 
The finite-frequency noise must therefore be derived from symmeterized
correlation functions, for details see 
\cite{Buttiker1992,Moskalets2004,Moskalets2014}. 
}

In order to analyze the physical content of the key result 
\eqref{IIIC_CurrCorrEx}, it is instructive to divide the noise power 
into two contributions, $P^{xy}_{\a\b}\equiv Q^{xy}_{\a\b}+
N^{xy}_{\a\b}$, that are given by $Q^{xy}_{\a\b}\equiv R^{xy}_{\a\b}
+ R^{yx}_{\a\b}$ and $N^{xy}_{\a\b}\equiv W^{xy}_{\a\b}+
C^{xy}_{\a\b}$ with\footnote{
To prove that the quantities $Q^{xy}_{\a\b}$ and $N^{xy}_{\a\b}$ 
indeed sum up to the total noise power \eqref{IIIC_CurrCorrEx}, use 
the sum rules \eqref{IIIB_SumRules}, the unitarity conditions
\eqref{IIB3_UniC} and shift the integration variables as needed.}
\begin{subequations}\label{IIIC_CCDecomp}
\begin{align}
\label{IIIC_CCDecompC}
R^{xy}_{\a\b}&\equiv\frac{1}{h}\sint{m}{E}\bigl(
	\d_{m0}\d_{\a\b}\zeta^x_E-|\FS^{\b\a+}_{m,E}|^2\zeta^x_{\Em}\bigr)
	\zeta^y_E f'^\b_E,\\[6pt]
\label{IIIC_CCDecompA}
W^{xy}_{\a\b}&\equiv\frac{1}{2h}\sint{m\g\d}{E}\;
	\Ac^{x\a,\g\d}_{m,E}\bar{\Ac}^{y\b,\g\d}_{m,E}
	(f^\g_E-f^\d_{\Em})^2,\\[6pt]
\label{IIIC_CCDecompB}
C^{xy}_{\a\b} &\equiv \d_{\a\b}\frac{1}{h}\sint{m\g}{E}\;
	|\FS^{\g\a+}_{m,E}|^2 \zeta^x_{\Em}\zeta^y_{\Em}
	(f'^\g_E-f'^\a_{\Em}).
\end{align}
\end{subequations}
Here, the thermal noise, or Nyquist-Johnson noise, $Q^{xy}_{\a\b}$,
results from thermal fluctuations in the incoming beams of carriers 
that emerge from the reservoirs. 
It remains finite in equilibrium but vanishes at zero temperature,
where thermal fluctuations are frozen out and $f'^{\a}_E =0$\footnote{
To be precise, we have $f'^\a_E\rightarrow 0$ for $E\neq\mu_\a$ and 
$f'^\a_E\rightarrow 1/4$ for $E=\mu_\a$ in the limit $T_\a\rightarrow
0$. 
Note that $f'^\a_E$ is the negative derivative of the Fermi function
$f^\a_E$ with respect to $(E-\mu_a)/T_\a$.}.
By contrast, the non-equilibrium noise $N^{xy}_{\a\b}$ vanishes 
if no external driving is applied to the conductor and all reservoirs
have the same temperature and chemical potential. 
Its first component, the shot noise $W^{xy}_{\a\b}$, which persists in
the zero-temperature limit, describes fluctuations in the matter and
energy currents due to the probabilistic nature of carrier 
transmissions and photon exchange between carriers and driving fields
in the quantum regime. 
Finally, the non-equilibrium correction, $C^{xy}_{\a\b}$, which 
vanishes at zero temperature, accounts for modulations of the thermal 
fluctuations in the outgoing beams of carriers due to thermochemical
biases and periodic driving. 

As a final remark for this section, we note that, although we have 
focused here on matter and energy currents, our analysis applies to
any set of generalized currents that can be represented by operators
of the form
\begin{equation}\label{IIIC_GenCurr}
\Jf'^{x}_{\a} = \sum\nolimits_{y\b} c^{xy}_{\a\b} \Jf^y_{\b}
\end{equation}
with real coefficients $c^{xy}_{\a\b}$. 
Specifically, the corresponding mean currents and the zero-frequency 
noise can be obtained directly from the formulas \eqref{IIIB_MC} and
\eqref{IIIC_CurrCorrEx} through the transformation rules
\begin{subequations}
\begin{align}
\label{IIIC_TransMC}
J'^x_\a &\equiv \limt \bigl\langle \Jf'^x_{\a,t}\bigr\rangle
	=\sum\nolimits_{y\b} c^{xy}_{\a\b} J^y_\b,\\[6pt]
\label{IIIC_TransCN}
P'^{xy}_{\a\b} &\equiv\limtt
	 \bigl\llangle \Jf'^x_{\a,t};\Jf'^y_{\b,t''}\bigr\rrangle\\[3pt]
&= \sum\nolimits_{u\g}\sum\nolimits_{v\d}
	c^{xu}_{\a\g}c^{yv}_{\b\d}P^{uv}_{\g\d},\nonumber
\end{align}
\end{subequations}
where $u=\rho,\E$, $v=\rho,\E$.
The thermal and quantum components of the transformed noise power, 
$P'^{xy}_{\a\b}\equiv Q'^{xy}_{\a\b}+N'^{xy}_{\a\b}\equiv 
R'^{xy}_{\a\b} + R'^{yx}_{\b\a}+ W'^{xy}_{\a\b}+ C'^{xy}_{\a\b}$, can
thus be identified by analogy as 
\begin{equation}\label{IIIC_TransCNComp}
A'^{xy}_{\a\b}= \sum\nolimits_{u\g}\sum\nolimits_{v\d}
	c^{xu}_{\a\g}c^{yv}_{\b\d}A^{uv}_{\g\d}
\end{equation}
for $A=Q,N,R,W,C$.

\section{Thermodynamics}\label{Sec_Thermo}
\subsection{The First Law}
The first law for periodically driven multi-terminal conductors 
follows directly from the conservation laws \eqref{IIIB_ConLaws} and 
can be formulated as 
\begin{align}\label{IVA_FirstL}
&\asum J^q_\a +\Pi_{{{\rm ac}}} - \Pi_{{{\rm el}}} = 0\\[6pt]
&\text{with}\quad J^q_\a\equiv J^\E_\a - \mu_\a J^\rho_\a 
	\quad\text{and}\quad
	\Pi_{{{\rm el}}} \equiv \asum (\mu-\mu_\a)J^\rho_\a, \nonumber
\end{align}
where $\mu$ denotes the base level of the chemical potential. 
It governs the balance between the thermal energy that is injected 
into the system by the reservoirs through the heat currents $J^q_\a$,
the mechanical power provided by the time dependent driving fields, 
$\Pi_{{{\rm ac}}}$, and the electrical power generated through the 
redistribution of carriers between the reservoirs, $\Pi_{{{\rm el}}}$.
Within the Floquet scattering approach, the first law 
\eqref{IVA_FirstL} is an immediate consequence of the sum rules 
\eqref{IIIB_SumRules}. 

\subsection{The Second Law}\label{Sec_SecondLaw}
The second law requires that the average rate of entropy production
that is caused by the transport process is non-negative, 
that is \cite{Callen1985}
\begin{equation}\label{IVA_SecondL}
\sigma \equiv -\asum J^q_\a/T_\a \geq 0. 
\end{equation}
A simple demonstration that the Floquet scattering approach is 
consistent with this constraint uses only the sum rules 
\eqref{IIIB_SumRules} and the fact that the Fermi distribution is 
the derivative of a convex function, for details see \cite{Nenciu2007,
Potanina2019a}.
In the following, we provide an alternative proof, which also shows 
that the dissipation rate $\sigma$ can only become zero if all
currents in the system vanish. 

Our proof is inspired by methods that are usually employed to derive 
bounds on quantum entropy functions, for details see \cite{Ohya1993}. 
The key idea is to express the rate of entropy production in terms of
the binary entropy function
\begin{equation}
\eta[a]\equiv -a\ln[a]-(1-a)\ln[1-a],
\end{equation}
and its first derivative, where $0\leq a\leq 1$. 
A quadratic lower bound on $\sigma$ can then be obtained from a simple
argument involving Taylor's theorem. 
We proceed as follows.
First, we use the formula \eqref{IIIB_MC} for the mean currents and 
the sum rules \eqref{IIIB_SumRules} to rewrite $\sigma$ as 
\begin{align}\label{IVA_EntAux1}
&h\sigma =\sint{m\a\b}{E}\; |\FS^{\b\a+}_{m,E}|^2
	\bigl((\Em-\mu_\a)/T_\a-(E-\mu_\b)/T_\b\bigr)f^\b_E
	\nonumber\\
&\begin{multlined}[t][1\displaywidth]
=\sint{m\a\b}{E} \; |\FS^{\b\a+}_{m,E}|^2\Bigl(
	\bigl(\ln[f^\b_E]-\ln[f^\a_{\Em}]\bigr)f^\b_E\\
		+\bigl(\ln[1-f^\b_E]-\ln[1-f^\a_{\Em}]\bigr)
		(1-f^\b_E)\Bigr)
\end{multlined}\nonumber\\
&=\sint{m\a\b}{E} \; |\FS^{\b\a+}_{m,E}|^2
	\bigl(\eta[f^\a_{\Em}]-\eta[f^\b_E] 
	+\eta_1[f^\a_{\Em}](f^\b_E-f^\a_{\Em})\bigr)
\end{align}
with $\eta_l[a]\equiv \partial^l_a\eta[a]$. 
By Taylor's theorem, there now exists a $g$ between $f^\b_E$ and
$f^\a_{\Em}$ such that 
\begin{align}
\label{IVA_EntAux2}
&\eta[f^\a_{\Em}]-\eta[f^\b_E] 
	+\eta_1[f^\a_{\Em}](f^\b_E-f^\a_{\Em})\\
&= -\eta_2[g](f^\b_E-f^\a_{\Em})^2/2.\nonumber
\end{align}
Since the Fermi function takes only values between $0$ and $1$, the 
number $g$ must also lie in this interval. 
Hence, we have $-\eta_2[g] = 1/g + 1/(1-g) \geq 4$ and therefore 
\begin{equation}\label{IVA_EntAux3}
\sigma\geq \frac{2}{h}\sint{m\a\b}{E} \; |\FS^{\b\a+}_{m,E}|^2
	\bigl(f^\a_{\Em}-f^\b_E\bigr)^2
\end{equation}
upon combining the Eqs.~\eqref{IVA_EntAux1} and \eqref{IVA_EntAux2}.

The bound \eqref{IVA_EntAux3} shows that, first, the rate of entropy
production can indeed not become negative within the Floquet 
scattering approach and, second, that $\sigma$ is zero if and only if
the integrand in Eq.~\eqref{IVA_EntAux3} vanishes for all energies
$E$ and all combinations of the indices $m,\a,\b$. 
Under this condition, however, all energy and particle currents must
also be zero according to Eq.~\eqref{IIIB_MCDiff}. 
We stress that this result, which was obtained here without any
assumptions on the behavior of the system under time reversal, should,
though intuitively expectable, not be regarded as trivial. 
In fact, the question whether or not dissipationless currents can
exist in normal conducting mesoscopic systems with broken time 
reversal symmetry has been the subject of an active debate in recent
years \cite{Benenti2017}. 

\subsection{Green-Kubo Relations}
The Green-Kubo relations are a cornerstone result of 
non-equilibrium statistical mechanics. 
They make it possible to express the linear response coefficients
that quantify the variations of mean currents due to a small changes
of the thermodynamic forces that drive the system away from
equilibrium in terms of integrated equilibrium correlation functions
of the involved currents \cite{Callen1985,Kubo1998}. 
As our final topic in this article, we will now show how this 
fundamental relationship arises naturally within the framework of
Floquet scattering theory. 

The thermodynamic forces, or affinities, for a transport process are
defined as gradients in the thermodynamic variables that form 
entropy-conjugate pairs with the conserved quantities of the system.
For a multi-terminal conductor, these objects can be identified with
the thermochemical biases between the external reservoirs, 
\begin{equation}\label{IVB_AffE}
F^\rho_\a \equiv \mu_\a/T_\a-\mu/T 
	\quad\text{and}\quad
	F^\E_\a \equiv 1/T-1/T_\a,
\end{equation}
where, $\mu$ and $T$ denote the base chemical potential and 
temperature.
Using these definitions, the rate of entropy production 
\eqref{IVA_SecondL} can be divided into a mechanical part,
$\sigma_{{{\rm ac}}}\equiv \Pi_{{{\rm ac}}}/T$, and a thermal one, 
$\sigma_{{{\rm th}}}\equiv \sigma -\sigma_{{{\rm ac}}}$, which now
assumes the characteristic bilinear form of irreversible 
thermodynamics \cite{Callen1985}, 
\begin{equation}\label{IVB_EntBiLin}
\sigma_{{{\rm th}}}= \asum F^\rho_\a J^\rho_\a +F^\E_{\a}J^\E_\a. 
\end{equation}
Several proposals were made to extend this structure to the total rate
of entropy production by associating the mechanical perturbation with
an effective current and a generalized affinity, which, depending on 
the scheme, corresponds to the mean applied work \cite{Izumida2010} or
either the amplitude \cite{Brandner2015f,Brandner2016,Proesmans2015a} 
or the frequency \cite{Ludovico2015b,Potanina2019a} of the periodic 
driving fields. 
For the purpose of our analysis, however, it is sufficient to focus on
the conventional thermal currents and affinities appearing in 
Eq.~\eqref{IVB_EntBiLin}. 

To establish the Green-Kubo relations for multi-terminal systems
we first calculate response coefficients 
\begin{equation}\label{IVB_RespCoeff}
L^{xy}_{\a\b} = \partial^y_\b J^x_\a
	= \frac{1}{h}\sint{m}{E}\;\bigl(\d_{m0}\d_{\a\b}\zeta^x_E 
	-|\FS^{\b\a+}_{m,E}|^2\zeta^x_{\Em}\bigr)\zeta^y_E f'^\beta_E,
\end{equation}
where we have used the current formula \eqref{IIIB_MC} and the symbol
$\partial^y_\b$ indicates the derivative with respect to the affinity
$F^y_\b$. 
Upon comparing this expression with the components of the current noise
given in the Eqs.~\eqref{IIIC_CCDecomp}, we find that $L^{xy}_{\a\b}
=R^{xy}_{\a\b}$ and thus 
\begin{equation}\label{IVB_FDT1}
L^{xy}_{\a\b} + L^{yx}_{\b\a} = Q^{xy}_{\a\b} 
	= P^{xy}_{\a\b}-N^{xy}_{\a\b}. 
\end{equation}
Hence, the symmetric part of the response coefficients 
\eqref{IVB_RespCoeff} is identical to the thermal noise, even if the
transport process takes place far from equilibrium. 
In equilibrium, i.e., for $F^x_\a =0$ and $V_t=0$, the non-equilibrium
noise $N^{xy}_{\a\b}$ vanishes and the relation \eqref{IVB_FDT1} 
becomes $(L^{xy}_{\a\b}+L^{yx}_{\b\a})|_{{{\rm eq}}}=
P^{xy}_{\a\b}|_{{{\rm eq}}}$. 
Moreover, provided that no magnetic fields are applied to the sample,
we recover the Onsager symmetry $L^{xy}_{\a\b}|_{{{\rm eq}}}
= L^{yx}_{\b\a}|_{{{\rm eq}}}$, as can be easily verified from the 
property \eqref{IIA2_AmplTRSym} of the stationary scattering 
amplitudes \cite{Callen1985}. 
We thus arrive at the standard form of the Green-Kubo relations for
multi-terminal conductors, 
\begin{equation}\label{IVB_GKEq}
2L^{xy}_{\a\b}|_{{{\rm eq}}} 
	=\limtt
	\bigl\llangle \Jf^x_{\a,t'};\Jf^y_{\b,t''}\bigr\rrangle
	\bigr|_{{{\rm eq}}}.
\end{equation}

In order to extend the result \eqref{IVB_GKEq} to non-equilibrium 
situations and systems with broken time-reversal symmetry, we have
to express the coefficient $L^{xy}_{\a\b}$ as an integrated 
correlation function that involves the current operator $\Jf^x_{\a}$. 
That is, we look for an observable $\If^x_{\a}$ that fulfills
\begin{equation}\label{IVB_GKGen}
L^{xy}_{\a\b} = \limtt \bigl\llangle \Jf^x_{\a,t};\If^y_{\b,t''}
	\bigr\rrangle. 
\end{equation}
A minimal choice for such a variable is given by 
\begin{equation}\label{IVB_Influx}
\If^x_\a \equiv \Eint\Epint
	\imath^{x\a}_{EE'} \sPhi^{\a\dagger}_{E,t}\sPhi^{\a}_{E',t},
\end{equation}
where the energy dependent weights, 
\begin{subequations}
\begin{align}
\imath^{\rho\a}_{EE'} &\equiv 
	\frac{1}{2h}[k_E k_{E'}]^{-\frac{1}{2}}(k_E+k_{E'}),\\[6pt]
\imath^{\E\a}_{EE'}   &\equiv 
	\frac{1}{4h}[k_E k_{E'}]^{-\frac{1}{2}}(k_E+k_{E'})(E+E'),
\end{align}
\end{subequations}
are found by replacing the scattering wave functions 
$\phi^{\a+}_{E,t}[r]$ in \eqref{IIIA_PaCurrME} and
\eqref{IIIA_EnCurrME} with the plane waves $w^-_E[r]$; 
recall Eq.~\eqref{IIA1_PWaves} for the definition of 
$w^-_E[r]$ and $k_E$. 
This operator can be easily shown to satisfy the condition 
\eqref{IVB_GKGen} by following the lines of Sec.~\ref{Sec_CN}. 
It describes the gross influx of matter $(x=\rho)$ or energy $(x=\E)$
from the reservoir $\a$ and thus provides a physically transparent
non-equilibrium generalization of the Green-Kubo relation 
\eqref{IVB_GKEq}, which covers even systems with broken time reversal
symmetry. 
From a practical perspective, the result \eqref{IVB_GKGen} makes it
possible to infer the time-integrated correlation function between net
currents and gross influx, which are otherwise hard to access, by 
measuring the variations of mean currents in response to small 
changes of the thermochemical biases \eqref{IVB_AffE}. 

We conclude this section by pointing out that the bilinear 
decomposition \eqref{IVB_EntBiLin} of $\sigma_{{{\rm th}}}$ into
affinities and currents is not unique. 
In fact, for any set of generalized currents and affinities,
\begin{align}
&J'^x_\a=\sum\nolimits_{y\b}=c^{xy}_{\a\b} J^y_\b
	\quad\text{and}\quad
	F'^x_\a=\sum\nolimits_{y\b}d^{xy}_{\a\b} F^y_\b\\[6pt]
&\text{with}\quad
	\sum\nolimits_{u\g} c^{ux}_{\g\a}d^{uy}_{\g\b}=\d_{xy}\d_{\a\b},
	\nonumber
\end{align}
the thermal rate of entropy production assumes the standard form
$\sigma_{{{\rm th}}} =\sum_{x\a} J'^x_\a F'^x_\a$. 
In particular, for the specific choice $c^{\rho x}_{\a\b}=
\d_{\a\b}\d_{x\rho}$ and $c^{\E x}_{\a\b} = 
\d_{\a\b}(\d_{x\E}-\mu_\a\d_{x\rho})$, the energy currents are
replaced by the heat current; 
that is, we have $J'^\E_\a = J^\E_\a-\mu_\a J^\rho_\a=J^q_\a$ and 
$F'^\E_\a = (\mu_\a-\mu)/T\equiv F^q_\a$. 
The Green-Kubo relations \eqref{IVB_GKEq} and their generalized 
counterparts \eqref{IVB_GKGen} are invariant under such linear 
transformations provided that the generalized influx operators are 
identified as $\If'^x_\a = \sum_{y\b}c^{xy}_{\a\b}\If^x_\a$. 
This result follows from the fact that the response coefficients
\eqref{IVB_RespCoeff} obey the same transformation rules as the 
zero-frequency current noise and its components, which are given
in the Eqs.~\eqref{IIIC_TransCN} and \eqref{IIIC_TransCNComp}. 
Specifically, we have 
\begin{equation}
L'^{xy}_{\a\b} \equiv \partial'^y_\b J'^x_\a 
	=\sum\nolimits_{u\g}\sum\nolimits_{v\d} 
	c^{xu}_{\a\g}c^{yu}_{\b\d} L^{uv}_{\g\d}
\end{equation}
as can be easily verified by inspection. 

\section{Perspectives and Challenges}\label{Sec_Open}
\subsection{Adiabatic Perturbation Theory}
In Sec.~\ref{Sec_PertTh}, we have shown how the Floquet scattering 
amplitudes can be calculated order by order in the dynamical
potential.
This approach is well justified if the periodic variations of the 
scattering potential are small compared to the typical carrier 
energies. 
For practical purposes, however, an adiabatic perturbation scheme,
where the frequency rather than the amplitude of the driving fields
plays the role of the expansion parameter, is often more suitable.

Such a theory can be developed as follows. 
Consider an approaching or escaping carrier with energy $E$ in the
terminal $\a$. 
If the dynamical potential is practically constant during the dwell 
time of this carrier inside the sample, its transition through the 
system at the time $t$ is described by the frozen scattering states 
$\ket{\kappa^{\a\pm}_{E,t}}$ \cite{Gasparian1996,Texier2015}. 
These states are solutions of the stationary Schr\"odinger equation 
\begin{equation}
H_t \ket{\kappa^{\a\pm}_{E,t}}= E\ket{\kappa^{\a\pm}_{E,t}}
\end{equation}
and satisfy the boundary conditions 
\begin{equation}\label{VIIIA_SBC}
\braket{r_\b}{\kappa^{\a\pm}_E}\equiv \kappa^{\a\pm}_E[r_\b]
	=\d_{\a\b} w^{\mp}_E[r_\b] + S^{\a\b\pm}_{E,t}w^{\pm}_E[r_\b]
\end{equation}
with the frozen scattering amplitudes given by
\begin{equation}
\braket{\kappa^{\b\mp}_{E',t}}{\kappa^{\a\pm}_{E,t}}
	= S^{\a\b\pm}_{E,t} \d_{E-E'}. 
\end{equation}
The corresponding quasi-static Floquet scattering amplitudes are
the Fourier components of these objects, i.e., 
\begin{equation}\label{VIIIA_Pert0}
\FS^{\a\b\pm}_{0m,E} = \frac{1}{\tau}\tint S^{\a\b\pm}_{E,t} u^m_t.
\end{equation}
This relation follows by comparing Eq.~\eqref{VIIIA_SBC} with 
Eq.~\eqref{IIB4_SBC} and assuming that the carrier energy is 
practically constant during the transition through the sample. 

The result \eqref{VIIIA_SBC} can be interpreted as the zeroth 
order of an expansion of the Floquet scattering amplitudes in 
the photon energy $\hbar\omega$. 
The first-order term of this series can be determined from a
phenomenological ansatz of the form \cite{Moskalets2012}
\begin{equation}\label{VIIIA_Pert1}
\FS^{\a\b\pm}_{1m,E} = \frac{\hbar\omega}{\tau}
	\tint m\partial_E^{\phantom{\b}} S^{\a\b\pm}_{E,t} u^m_t 
	\pm\hbar\omega\mathcal{A}^{\a\b\pm}_{m,E}.
\end{equation} 
Here, the first term accounts for small changes in the carrier energy
during the transition and the correction $\mathcal{A}^{\a\b\pm}_{m,E}$
is chosen such that the approximated Floquet scattering 
amplitudes obey the unitarity conditions \eqref{IIB3_UniC}. 

This scheme proved quite effective for various practical applications
\cite{Moskalets2002,Moskalets2004}. 
How it can be derived from a systematic perturbation theory, which
would make it possible to calculate also higher-order terms, however,
is not immediately clear.
As a first attempt, we might try to emulate the adiabatic perturbation
theory for master equations with slowly varying parameters by adapting
the Lippmann-Schwinger formalism of Sec.~\ref{Sec_IIB2} 
\cite{Cavina2017a,Potanina2019}. 
To this end, the free scattering vectors \eqref{IIB2_FreeStates} have 
to be replaced with their frozen counterparts, 
\begin{equation}
\langle t|\kappa^{m\a\pm}_E\rrangle
	\equiv u^m_t\ket{\kappa^{\a\pm}_{\Em,t}}. 
\end{equation}
The roles of the free effective Hamiltonian and the perturbation are 
then assumed by the operators $\hat{K}$ and $\hat{D}$, respectively,
which are defined as $\bra{t}\hat{K}\Ket{\psi}\equiv H_t\ket{\psi_t}$
and  $\bra{t}\hat{D}\Ket{\psi}\equiv-i\hbar\partial_t\ket{\psi_t}$. 
Upon repeating the derivations of Sec.~\ref{Sec_IIB2}, we thus find 
that the Floquet scattering amplitudes, up to second-order
contributions in $\hat{D}$, read
\begin{align}
\label{VIIIA_PertLS}
\FS^{\a\b\pm}_{m,E} &= \frac{1}{\tau}\tint S^{\a\b\pm}_{\Em,t} u^m_t
	\mp 2\pi i\Bra{\kappa^{0\b\mp}_{\Em}}\hat{D}
	\Ket{\kappa^{m\a\pm}_E}\\[3pt]
&= \frac{1}{\tau}\tint S^{\a\b\pm}_{\Em,t} u^m_t
	\pm\hbar\omega\tint 
	\braket{\dot{\kappa}^{\b\mp}_{\Em,t}}{\kappa^{\a\pm}_{\Em,t}}u^m_t
	\nonumber\\[3pt]
&\begin{multlined}[t][.91\displaywidth]
	=\frac{1}{\tau}\tint S^{\a\b\pm}_{E,t} u^m_t
	+\frac{\hbar\omega}{\tau}\tint 
	m\partial_E^{\phantom{\beta}} S^{\a\b\pm}_{E,t} u^m_t\\[3pt]
	\pm\hbar\omega\tint 
	\braket{\dot{\kappa}^{\b\pm}_{E,t}}{\kappa^{\a\pm}_{E,t}}u^m_t
	+\mathcal{O}[(\hbar\omega)^2].
	\nonumber\end{multlined}
\end{align}
Hence, we indeed recover the zeroth- and fist-order terms
\eqref{VIIIA_Pert0} and \eqref{VIIIA_Pert1}. 
This result, however, must be taken with a grain of salt, since the 
correction term in Eq.~\eqref{VIIIA_PertLS}, which involves the time
derivative of the frozen scattering state 
$\ket{\kappa^{\b\pm}_{E,t}}$, is generally divergent. 
Therefore, the expression \eqref{VIIIA_PertLS} should not be regarded
as a proper expansion of the Floquet scattering amplitudes.

The singular behavior of the last term in Eq.~\eqref{VIIIA_PertLS} arises 
because the time derivative $\hat{D}$, in contrast to the dynamical 
potential $\hat{V}$, which vanishes outside the scattering region,
constitutes an unbounded operator on
$\hat{\mathcal{H}}$.
To overcome this problem, it might be necessary to invoke techniques
of singular perturbation theory, adiabatic gauge potentials
\cite{Kolodrubetz2017,Weinberg2017} or a transformation of the 
scattering amplitudes into the time domain;
the latter approach lead to a consistent first-order expansion in
\cite{Thomas2012}. 
We leave it as a challenge for future studies to derive a systematic
adiabatic perturbation theory by further developing the formalism 
presented in this article

\subsection{Thermal Machines}
The Floquet scattering formalism provides a universal platform to
explore the performance of thermal nano-devices.
As a concrete example, we might consider a quantum heat engine 
that consists of a driven sample and two reservoirs with equal 
chemical potential $\mu$ and different temperatures 
$T_1\equiv T_{{{\rm c}}}$ and $T_2\equiv T_{{{\rm h}}}>T_{{{\rm c}}}$.
Here, we imagine that the variations of the scattering potential
are caused by the motion of mechanical degrees of freedom like a 
mesoscopic paddle wheel, which perform work against some external load
\cite{Bustos-Marun2013,Arrachea2016,Bruch2018a}.
The thermodynamic performance of such a machine is determined by two
benchmark parameters, its mean power output $-\Pi_{{{\rm ac}}}$ and
its efficiency $\eta\equiv - \Pi_{{{\rm ac}}}/J^q_{{{\rm h}}}$. 
The latter figure is thereby subject the universal Carnot bound 
\begin{equation}
\eta \leq \eta_{{{\rm C}}}\equiv 1- T_{{{\rm c}}}/T_{{{\rm h}}},
\end{equation}
which follows from the second law, $\sigma=J^q_{{{\rm h}}}
F^q_{{{\rm h}}}+\Pi_{{{\rm ac}}}/T_{{{\rm c}}}\geq 0$, and can be
attained only in the quasi-static limit, where $-\Pi_{{{\rm ac}}}$
goes to zero\footnote{This result follows from the fact that the 
rate of entropy production $\sigma$ vanishes only if all currents 
are zero, see Sec.~\ref{Sec_SecondLaw}.}.

From a practical perspective, it is therefore important to determine 
the maximum efficiency, at which a nano-engine can deliver a given 
power output. 
For autonomous, i.e., thermoelectric, heat engines such bounds have 
been found by seeking constraints on the total rate of entropy
production that go beyond the second law \cite{Whitney2013,
Brandner2015,Shiraishi2016b,Pietzonka2017,Shiraishi2018b}, or by 
explicitly optimizing the scattering amplitudes of the sample 
\cite{Whitney2014,Whitney2014a,Hofer2015a,Sanchez2015a,
Samuelsson2017}.
The first strategy has also been applied in studies of piston-type 
heat engines, which use a closed working system, and lead to the 
general trade-off relation
\begin{equation}
\eta(\eta_{{{\rm C}}}-\eta) \geq - \Pi/\Theta
\end{equation}
between efficiency $\eta$ and power output $-\Pi$; 
here, $\Theta > 0$ is a system-specific constant \cite{Brandner2015f,
Brandner2016,Shiraishi2016b,Shiraishi2018b}. 
First steps towards an extension of this bound to paddle-wheel 
type quantum engines, which are driven by a continuous flow of 
carriers, have been made under the assumptions of slowly varying 
driving fields and small thermochemical biases 
\cite{Ludovico2015b,Potanina2019a}. 
A universal and physically transparent performance bound that covers
also devices operating far from  equilibrium is, however, still
lacking. 

\subsection{Thermodynamic Uncertainty Relations}
Thermodynamic uncertainty relations describe a trade-off between 
dissipation and precision in stationary non-equilibrium processes. 
Specifically, for a time-homogeneous Markov process that obeys 
detailed balance, the inequality 
\begin{equation}\label{VIIIC_TUR}
\sigma\epsilon^2 \geq 2 \quad\text{with}\quad 
	\epsilon\equiv \sqrt{2P/J^2}
\end{equation}
holds for arbitrary currents with mean value $J$ and fluctuations, or
noise power, $P$, where $\sigma$ denotes the total rate of entropy
production and $\epsilon$ the relative uncertainty of the current $J$
\cite{Barato2015,Gingrich2016}. 
This bound, which was first discovered for biomolecular processes,
does, however, not apply to periodically driven systems, systems with
broken time-reversal symmetry or in the quantum regime
\cite{Barato2016,Brandner2017b,Ptaszynski2018,Holubec2018}. 
In order to close these gaps, a whole variety of generalized
thermodynamic uncertainty relations have been proposed over the last
years, see for instance \cite{Proesmans2017b,Barato2018b,Koyuk2019b,
Macieszczak2018,Koyuk2019,Carollo2019,Hasegawa2019a}.

A particularly transparent result was recently obtained in 
\cite{Koyuk2019b}, where the frequency dependent bound 
\begin{equation}
\sigma_\omega^{\phantom{2}}\epsilon_\omega^2
	\geq 2[1- \omega (\partial_\omega J_\omega)/J_\omega]^2
\end{equation}
was derived for periodically driven Markov jump processes. 
Whether or not this result can be extended to coherent mesoscopic
conductors, or whether the relation \eqref{VIIIC_TUR} can be 
generalized for such systems by other means are compelling questions,
which can be systematically investigated within the theoretical 
framework presented in this article.
Further research in this direction promises valuable insights on how
quantum effects can be exploited to control the thermodynamic cost of
precision in transport processes. 
However, this endeavor can be expected to be challenging, since 
general properties of the Floquet scattering amplitudes that go beyond
the ones discussed in Sec.~\ref{Sec_FST_DC} are hard to establish and 
specific models for which they can be determined exactly are scarce. 

\newcommand{\kint}{\int_0^\infty \!\!\! dk \;}
\newtheorem{lemma}{Lemma}
\appendix 
\section{Some Helpful Lemmas}\label{ApxA}
\noindent {{\bf\sffamily Lemma 1a.}}
Let $F^\pm_z$ be a complex function that is bounded and holomorphic on
the stripe $D^\pm \equiv [0,\infty)\times [0,\pm iR]$ with $R>0$. 
Then, for any $v>0$, we have 
\begin{subequations}
\begin{align}
\label{Lem1_1a}
&\lim_{\ve\rightarrow 0}\int_0^\infty \!\!\!du 
	\frac{\exp[\pm i x u]}{u^2-v^2\mp i\ve} F_u^\pm
	\asymp \pm \frac{\pi i \exp[\pm i x v]}{v} F_v^\pm,\\
\label{Lem1_1b}
&\lim_{\ve\rightarrow 0}\int_0^\infty \!\!\!du 
	\frac{\exp[\pm i x u]}{u^2+v^2\mp i\ve} F_u^\pm
	\asymp 0
\end{align}
\end{subequations}
in the limit $x\rightarrow\infty$. 

\begin{figure}
\begin{center}
\includegraphics[scale=1.055]{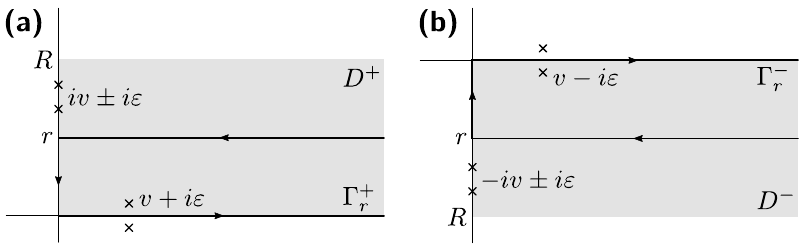}
\end{center}
\caption{
Graphical illustration of the integration contours used in 
the proof of \emph{Lemma 1}. 
(a) The contour $\Gamma^+_r$ encircles a rectangle with height $r$ and
infinite width, whose lower edge falls on the positive real axis. 
Crosses indicate the singularities of the integrands in the 
Eqs.~\eqref{Lem1_2} and \eqref{Lem1_3}. 
(b) The same picture for the contour $\Gamma^-_r$. 
\label{Fig_3}}
\end{figure}

\noindent {{\bf\sffamily Proof.}}
We proceed in two steps. 
First, we close the integration path in the complex plane as shown in
Fig.~\ref{Fig_3} and observe that 
\begin{align}\label{Lem1_4}
&\int_{\Gamma^\pm_r} dz
	\frac{\exp[\pm i x z]}{z^2-v^2\mp i\ve} F_z^\pm
	\asymp \int_0^\infty \!\!\! du
	\frac{\exp[\pm i x u]}{u^2-v^2\mp i\ve} F_u^\pm,\\
&\int_{\Gamma^\pm_r} dz
	\frac{\exp[\pm i x z]}{z^2+v^2\mp i\ve} F_z^\pm
	\asymp \int_0^\infty \!\!\! du
	\frac{\exp[\pm i x u]}{u^2+v^2\mp i\ve} F_u^\pm
	\nonumber
\end{align}
for $x\rightarrow\infty$, since the integrand on the left-hand side is
exponentially suppressed in $x$ on either the upper ($+$) or the lower 
($-$) half plane. 
Second, using Cauchy's theorem to evaluate the contour integral yields
\begin{subequations}
\begin{align}
\label{Lem1_2}
& \lim_{\ve\rightarrow 0}\int_{\Gamma^\pm_R} dz
	\frac{\exp[\pm i x z]}{z^2-v^2\mp i\ve} F_z^\pm\\
&= \lim_{\ve\rightarrow 0} \int_{\Gamma^\pm_R} dz
	\left[\frac{1}{z-v \mp i\ve}-\frac{1}{z+v\pm i\ve}\right]
	\frac{\exp[\pm i xz]}{2v}F^\pm_z \nonumber\\
&= \pm \frac{\pi i \exp[\pm i x v]}{v} F_v^\pm 
\quad\text{and}\nonumber\\[3pt]
\label{Lem1_3}
&\lim_{\ve\rightarrow 0}\int_{\Gamma^\pm_r} dz
	\frac{\exp[\pm i x z]}{z^2+v^2\mp i\ve} F_z^\pm\\
&= \lim_{\ve\rightarrow 0} \int_{\Gamma^\pm_r} dz
	\left[\frac{1}{z-iv\mp i\ve}-\frac{1}{z+iv\pm i\ve}\right]
	\frac{\exp[\pm i xz]}{2v}F^\pm_z =0,\nonumber
\end{align}
\end{subequations}
where we set $r=R$ in Eq.~\eqref{Lem1_2} and $0<r<v$ in 
Eq.~\eqref{Lem1_3}.

\vspace*{6pt}
\noindent {{\bf\sffamily Lemma 1b.}}
For $F^\pm_z$ as in Lemma~1a and $w\neq 0$ being real, we have 
\begin{equation}
\lim_{\ve\rightarrow 0} \int_0^\infty \!\!\! du 
	\frac{\exp[\mp ix u]}{u^2 + w \mp i\ve}F^\pm_u\asymp 0
\end{equation}
in the limit $x\rightarrow\infty$. 

\noindent {{\bf\sffamily Proof.}}
Set $w=-v^2$ for $w<0$ and $w=v^2$ for $w>0$ and repeat the steps of 
the proof of Lemma~1.

\vspace*{6pt}
\noindent {{\bf\sffamily Lemma 1c.}}
For $F_z^\pm$ as in Lemma~1a, $G^\pm_z\equiv F^\pm_{\sqrt{z}}$ and 
$w\neq 0$ being real, we have 
\begin{subequations}
\begin{align}
&\lim_{\ve\rightarrow 0} \int_0^\infty \!\!\! du 
	\frac{\exp[\pm ix \sqrt{u}]}{u-w\mp i\ve} \frac{G^\pm_u}{\sqrt{u}}
	\asymp \pm 2\pi i \frac{\exp[\pm i x \sqrt{w}] G^\pm_w}{\sqrt{w}},\\
&\lim_{\ve\rightarrow 0} \int_0^\infty \!\!\! du 
	\frac{\exp[\mp ix \sqrt{u}]}{u-w\mp i\ve} \frac{G^\pm_u}{\sqrt{u}}
	\asymp 0
\end{align}
in the limit $x\rightarrow \infty$.
\end{subequations}

\noindent {{\bf\sffamily Proof.}}
Change the integration variable to $s\equiv \sqrt{u}$ and apply the
Lemmas~1a and 1b.  

\vspace*{6pt}
\noindent {{\bf\sffamily Lemma 2.}}
Let $F_u$ be a test function on the real axis and define $\theta_u
\equiv (1-\exp[-iu])/(iu)$.
Then, for any integers $m$ and $n$, we have 
\begin{equation}\label{Lem2_1}
\lim_{x\rightarrow\infty} \int\!\!\!du\; x\cdot
	\theta_{(u-m)x}\bar{\theta}_{(u-n)x} F_u
	= 2\pi \d_{mn} F_{m}. 
\end{equation}

\noindent {{\bf\sffamily Proof.}}
We first rewrite the left-hand side of Eq.~\eqref{Lem2_1} as  
\begin{align}
& \lim_{x\rightarrow\infty}\int\!\!\! du \; x\cdot
	|\theta_{(u-m)x}|^2 \frac{\bar{\theta}_{(u-n)x}}{
	\bar{\theta}_{(u-m)x}}F_u\\
&= \lim_{x\rightarrow\infty}\int\!\!\! du \; x\cdot
	{{{\rm sinc}}}^2[(u-m)x/2]\frac{\bar{\theta}_{(u-n)x}}{
	\bar{\theta}_{(u-m)x}}F_u,
	\nonumber
\end{align}
where ${{{\rm sinc}}}[u]\equiv \sin[u]/u$. 
Next, we observe that the function ${{{\rm sinc}}}^2[u]$ assumes only 
non-negative values and obeys 
\begin{subequations}
\begin{align}
&\lim_{x\rightarrow\infty}\int_{|u|\leq \ve} \!\!\! du\;
	 x\cdot{{{\rm sinc}}}^2[ux] = \pi 
	 \quad\text{and}\\[6pt]
&\lim_{x\rightarrow\infty} x\cdot{{{\rm sinc}}}^2[ux]=0
	\quad\text{for}\quad \ve < |u| <1/\ve,
\end{align}
\end{subequations}
where $\ve>0$. 
Consequently, we have \cite{Appel2007}
\begin{align}
& \lim_{x\rightarrow\infty}\int\!\!\! du \; x\cdot
	{{{\rm sinc}}}^2[(u-m)x/2] \frac{\bar{\theta}_{(u-n)x}}{
	\bar{\theta}_{(u-m)x}}F_u\\
& = 2\pi  \lim_{x\rightarrow\infty}
	\int \!\!\! du \; \bar{\theta}_{(m-n)x}F_{m} 
	= 2\pi\d_{mn}F_{m},\nonumber
\end{align}
where we used that $\lim_{u\rightarrow 0}\theta_u=1$ and 
$\lim_{x\rightarrow\infty}\theta_{mx}=0$ for any $m\neq 0$. 

\begin{acknowledgments}
The author acknowledges insightful discussions with E. Potanina, 
M. Moskalets, K. Saito and U. Seifert.
The research leading to the results presented in this article has 
received funding from the Academy of Finland (Contract No. 296073),
the Japan Society for the Promotion of Science through a Postdoctoral
Fellowship for Research in Japan (Fellowship ID: P19026), the 
University of Nottingham through a Nottingham Research Fellowship and
from UK Research and Innovation through a Future Leaders Fellowship
(Grant Reference: MR/S034714/1).
\end{acknowledgments}

%

\end{document}